\documentclass[11pt]{article}
\usepackage[margin=1in]{geometry}
\usepackage{graphicx}
\usepackage{amsmath}
\usepackage{hyperref}
\usepackage{cite}
\usepackage{url}
\usepackage{algorithm}
\usepackage{algpseudocode}
\usepackage{amsfonts}
\usepackage{subcaption}
\usepackage{siunitx}
\usepackage{booktabs}
\usepackage{multirow}
\usepackage{longtable}
\usepackage{placeins}

\title{\LARGE \bf 
Simulating Evolvability as a Learning Algorithm: Empirical Investigations on Distribution Sensitivity, Robustness, and Constraint Tradeoffs\thanks{Code available at: \url{https://github.com/fiidalgo/boolean-evolvability}}}

\author{
  \normalsize Nico Fidalgo \quad Puyuan Ye \\
  \normalsize Harvard University \\
  \normalsize \texttt{nfidalgo@college.harvard.edu, puyuanye@college.harvard.edu}
}

\date{May 14, 2025}

\begin{document}

\maketitle

\begin{abstract}
The theory of evolvability, introduced by Valiant (2009), formalizes evolution as a constrained learning algorithm operating without labeled examples or structural knowledge. While theoretical work has established the evolvability of specific function classes under idealized conditions, the framework remains largely untested empirically. In this paper, we implement a genetic algorithm that faithfully simulates Valiant’s model and conduct extensive experiments across six Boolean function classes: monotone conjunctions, monotone disjunctions, parity, majority, general conjunctions, and general disjunctions. Our study examines evolvability under uniform and non-uniform distributions, investigates the effects of fixed initial hypotheses and the removal of neutral mutations, and highlights how these constraints alter convergence behavior.

We validate known results (e.g., evolvability of monotone conjunctions, non-evolvability of parity) and offer the first empirical evidence on the evolvability of majority and general Boolean classes. Our findings reveal sharp performance drops at intermediate dimensions (e.g., \(n=10\)) and expose the essential role of neutral mutations in escaping fitness plateaus. We also demonstrate that evolvability can depend strongly on the input distribution. These insights clarify practical limits of evolutionary search and suggest new directions for theoretical work, including potential refinements to evolvability definitions and bounds. Our implementation provides a rigorous, extensible framework for empirical analysis and serves as a testbed for future explorations of learning through evolution.
\end{abstract}

\section{Introduction}

Evolution is nature’s algorithm for solving problems under uncertainty. From the emergence of biochemical pathways to the development of the human brain, evolution has enabled the discovery of complex, adaptive mechanisms without explicit design or foresight. But can evolution itself be modeled as an algorithm for learning? And if so, under what conditions does it work? These questions form the foundation of Valiant’s groundbreaking 2009 paper, which introduced the theory of evolvability---a formal framework to analyze the power and limitations of evolutionary processes through the lens of computational learning theory~\cite{valiant2009evolvability}.

At its core, evolvability models how a population of candidate hypotheses can evolve toward an ideal function through local, performance-guided mutations, with no access to labeled examples or syntactic structure---only aggregate fitness values over a distribution of environmental conditions. Valiant shows that evolvability is strictly weaker than both PAC learnability and the Statistical Query (SQ) model, highlighting the constraints faced by biological evolution and its algorithmic analogs. For instance, classes like monotone conjunctions and disjunctions are shown to be evolvable under the uniform distribution, while parity functions, though PAC-learnable, are not evolvable due to their poor statistical query performance.

This theory is not merely theoretical abstraction—it provides a rigorous framework to examine which mechanisms can arise through evolution under realistic constraints on time, population size, and available feedback. The evolvability model forces us to ask not just whether a class is learnable, but whether it can be discovered gradually and blindly by a population navigating a noisy fitness landscape with limited memory and no interpretability. These constraints reflect real limits in biological systems and present a serious challenge for artificial ones. In this sense, evolvability serves as both a biological metaphor and a rigorous testbed for studying learning under constraints that mirror the real world.

Despite its elegance, the theory of evolvability remains largely underexplored outside the realm of formal proofs. Existing work has mostly focused on characterizing a few special function classes theoretically and has seldom addressed the broader empirical behavior of the model. Evolvability results are often proven for specific distributions (typically uniform), under idealized mutation operators, and for narrow hypothesis spaces. This leaves major questions open: Are these results robust to shifts in distribution or modeling assumptions? Do they generalize to other natural function classes like majority or general Boolean formulas? Can meaningful learning still occur when key assumptions, such as random initialization or neutral mutations, are removed?

Our project addresses these questions by building an empirical framework that simulates Valiant's model faithfully while pushing it into new experimental regimes. We implement a genetic algorithm (GA)-based simulation of the evolvability model and systematically explore its behavior along three axes. First, we empirically test the evolvability of several Boolean function classes. These include monotone conjunctions, disjunctions, and parity functions, which serve as theoretical baselines, as well as majority functions, general conjunctions, and general disjunctions, which lack formal evolvability characterizations. Second, we evaluate distributional robustness by testing evolvability under multiple input distributions, addressing Valiant’s suggestion that functions which evolve regardless of distribution are particularly valuable. Third, we test the effect of constrained modeling assumptions—such as fixed initial hypotheses and the disallowance of neutral mutations—on convergence and overall performance. These experiments draw directly from open questions raised in the concluding discussion of Valiant’s original paper.

This work is not intended to provide formal proofs, but rather to offer experimental insights that can inform and inspire theoretical analysis. For function classes such as majority or general conjunctions, our results provide the first empirical evidence of their behavior in this model. These findings can motivate new directions in formal learning theory to investigate whether the observed performance can be guaranteed under realistic conditions. Moreover, by showing which classes are sensitive to distributional shift or model constraints, we clarify where evolvability’s power lies and where its limitations emerge. Our implementation also provides a testbed for future empirical studies, including ones that might leverage larger-scale computation, alternative mutation operators, or continuous representations.

More broadly, our project contributes to a growing conversation in theoretical computer science about the boundaries of feasible learning. In an age of increasingly complex machine learning models, the evolvability framework reminds us that meaningful behavior must often be discovered incrementally, blindly, and with limited feedback. Understanding what can evolve under these constraints—both in nature and in artificial systems—remains a fundamental challenge. We hope that this work not only validates and extends existing theory, but also stimulates deeper investigation into what makes learning possible in the most restricted and biologically plausible settings.

\section{Related Work}

Valiant’s theory of evolvability~\cite{valiant2009evolvability} builds on the rich tradition of computational learning theory, particularly PAC learning~\cite{valiant1984theory} and the Statistical Query model~\cite{kearns1998efficient}. It introduces a restricted model where updates to hypotheses can only depend on their aggregate performance with respect to a distribution, mimicking the way natural selection operates on organisms’ phenotypic fitness without explicit understanding of genotypic structure.

Within this framework, Valiant proved that monotone conjunctions and disjunctions are evolvable under the uniform distribution, while parity functions are not. These results provided the first rigorous delineation of which classes can evolve under limited feedback and feasible resources. Subsequent work by Feldman~\cite{feldman2008evolvability} connected evolvability more deeply to the Statistical Query model, showing that evolvability implies learnability via statistical queries, and introduced more general criteria for evolvability.

However, much of the existing literature remains theoretical. There are few empirical studies that test whether function classes provably evolvable in theory also evolve in practice under simulated evolutionary conditions. One exception is the work by Ros and Sollich~\cite{ros1997neural}, who explored learning conjunctions and disjunctions through evolutionary algorithms but used access to labeled examples and relied on input-specific information, thereby departing from the constraints of the evolvability model. Similarly, research in neuroevolution and evolutionary computation has explored evolving neural architectures~\cite{stanley2002evolving}, but often with fitness feedback based on specific instance-level evaluation, violating the aggregate-only criterion essential to Valiant’s framework.

In addition to these algorithmic studies, recent theoretical work has attempted to analyze the evolvability of function classes such as majority, threshold, and general disjunctions. Blum et al.~\cite{blum2005practical} examined learnability of threshold functions under various constraints, while recent advances in robust agnostic learning suggest that majority-like functions might be learnable in noisy settings~\cite{feldman2010agnostic}, although no conclusive evolvability proofs are available. To our knowledge, there has been no empirical analysis testing the evolvability of majority functions or general conjunctions/disjunctions under Valiant's framework.

Our work contributes to this gap in three primary ways. First, we empirically validate known theoretical results using a concrete simulation of Valiant’s model. Second, we explore new function classes (majority, general conjunctions, general disjunctions) for which evolvability has not been formally established. Third, we test evolvability under more realistic, distributionally varied and constrained settings, directly inspired by the open questions raised in Valiant’s original discussion. Specifically, we address whether evolvability generalizes across input distributions---a key desideratum for biological realism and algorithmic robustness---and also examine how relaxing or tightening the framework (e.g., fixed starting points, banning neutral mutations) affects the evolutionary process.

By grounding our methodology in the formal theory and extending it with rigorous simulation, our work seeks to operationalize evolvability as a viable model of learning, contributing both empirical validation and new insight into its boundaries and behavior.

\section{Background}

The theory of evolvability, introduced by Valiant~\cite{valiant2009evolvability}, offers a formalization of biological evolution within computational learning theory. It models the evolutionary process as a type of learning algorithm constrained by biological limitations: limited memory, no explicit access to labeled examples, and reliance solely on aggregate feedback from the environment. This makes evolvability strictly more constrained than traditional learning models such as PAC (Probably Approximately Correct) learning or the Statistical Query (SQ) model. In particular, updates to hypotheses in the evolvability model must be driven by their empirical performance—an aggregate measure of how well the candidate hypothesis agrees with the target function over a distribution—rather than by syntactic or structural insights derived from specific inputs.

In this framework, let $X_n = \{ -1, 1 \}^n$ be the space of all possible inputs, where each input vector $x$ encodes an environmental state composed of $n$ binary variables. A function $f : X_n \rightarrow \{-1, 1\}$ represents the ideal target behavior in response to these environmental conditions. The goal of the evolutionary process is to find a hypothesis $r : X_n \rightarrow \{-1, 1\}$ such that $r$ closely approximates $f$. The performance of a representation $r$ with respect to $f$ under a distribution $D_n$ is defined as:

\[
\text{Perf}_f(r, D_n) = \sum_{x \in X_n} f(x) r(x) D_n(x)
\]

This performance metric ranges from $-1$ to $1$, where a value of $1$ indicates perfect agreement between $r$ and $f$ on all inputs with non-zero probability under $D_n$. Since the true distribution is rarely accessible in full, empirical performance is used instead. Given a sample set $Y = \{ x_1, \ldots, x_s \}$ drawn independently from $D_n$, empirical performance is calculated as:

\[
\text{Perf}_f(r, D_n, s) = \frac{1}{s} \sum_{i=1}^s f(x_i) r(x_i)
\]

This empirical performance informs the mutator, a procedure that selects the next generation's hypothesis from a neighborhood of candidates. For each representation $r$, the neighborhood $N(r, \epsilon)$ is a set of size at most $p(n, 1/\epsilon)$, generated by a randomized process. Candidates in the neighborhood are evaluated based on their empirical performance. Let $v(r) = \text{Perf}_f(r, D_n, s)$ denote the performance of a representation. Then we define the set of beneficial mutations as:

\[
\text{Bene} = \{ r' \in N(r, \epsilon) \mid v(r') \ge v(r) + t \}
\]

and the set of neutral mutations as:

\[
\text{Neut} = \{ r' \in N(r, \epsilon) \mid v(r') \ge v(r) - t \} \setminus \text{Bene}
\]

If there exists any $r' \in \text{Bene}$, the mutator selects one uniformly at random (or via a distribution that respects generation probability). If $\text{Bene}$ is empty, the mutator selects from $\text{Neut}$. This mechanism reflects natural selection: improvements are preferred, but neutral variations are permitted when no improvements exist.

A class of functions $C$ is said to be evolvable by a representation class $R$ over a distribution $D$ if, starting from any initial hypothesis $r_0 \in R$ and given tolerance $\epsilon > 0$, the sequence of hypotheses produced by successive applications of the mutator yields, with high probability, a final hypothesis $r$ satisfying:

\[
\text{Perf}_f(r, D_n) > 1 - \epsilon
\]

within at most $g(n, 1/\epsilon)$ generations. Our implementation closely adheres to this formalism by defining Boolean function classes as target functions, generating empirical distributions, evaluating fitness empirically, and mutating candidate hypotheses within bounded neighborhoods consistent with these definitions.

While Valiant’s original paper formally establishes the evolvability of monotone conjunctions and disjunctions under the uniform distribution, and rules out the evolvability of parity functions under the same conditions, many questions remain open regarding other function classes and the practical behavior of the model. One particularly relevant gap in the literature is the lack of formal or empirical study of the evolvability of threshold-based functions such as majority. Although majority functions are conceptually simple and biologically plausible, they are not known to be evolvable in the original framework, and no prior work has demonstrated empirically whether local mutations guided by aggregate fitness can reliably converge to them. The majority function is known to be PAC-learnable using a variety of techniques, including boosting and margin-based algorithms~\cite{blum2005practical}, but it has not been shown to be evolvable under Valiant’s constraints.

A similar gap exists for general conjunctions and disjunctions. While the evolvability of their monotone variants is provable under the uniform distribution, allowing for negated literals dramatically increases the complexity of the hypothesis space. Feldman~\cite{feldman2008evolvability} and others have analyzed how the statistical query model relates to evolvability and highlighted how subtle differences in the learning assumptions—such as representation complexity or access to performance signals—can change which function classes are evolvable. However, no known work formally proves or refutes the evolvability of general conjunctions or disjunctions, and empirical evidence is similarly scarce.

Although a few related efforts in evolutionary computation and neuroevolution have explored function learning through biologically inspired algorithms~\cite{ros1997neural, stanley2002evolving}, these approaches generally allow feedback from labeled examples or incorporate structural information that Valiant’s framework explicitly disallows. Consequently, they do not address the same constraints or offer conclusive insight into what is evolvable in the theoretical sense. Moreover, most existing empirical work on evolution-inspired learning either evaluates performance on engineering tasks (e.g., neural architecture search) or bypasses theoretical constraints by using heuristics or problem-specific encodings. As a result, Valiant’s model remains underexplored from an empirical standpoint, especially with respect to its core question: which function classes can evolve with only aggregate feedback and modest computational resources?

This project was motivated in large part by this gap. We aimed to test the evolvability of majority, general conjunctions, and general disjunctions—classes that have not been theoretically settled—and to do so using an empirical implementation that faithfully adheres to the constraints of the original framework. In doing so, we also sought to verify known results for benchmark classes like monotone conjunctions and parity, providing a sanity check for our implementation while extending the boundaries of what is known empirically about evolvability in Boolean function learning.

\section{Experimental Methodology}

We implemented the evolvability framework using a genetic algorithm (GA) coded in Python. The primary goal of our simulation was to empirically assess the evolvability of several Boolean function classes, including both those that are provably evolvable in theory, such as monotone conjunctions and disjunctions, and those that are less well-understood or even provably non-evolvable, such as parity. Additionally, we investigated classes such as majority and general conjunction/disjunction, for which evolvability has not been theoretically established.

Each Boolean function class is instantiated from a randomly sampled support set $S \subseteq \{x_1, \ldots, x_n\}$, where $n$ is the input dimension. For monotone conjunctions, the function evaluates to true if all variables in $S$ are true. For example, the function $x_1 \land x_3 \land x_7$ returns true only when each of $x_1$, $x_3$, and $x_7$ equals $+1$. Monotone disjunctions return true if at least one variable in $S$ is true. Parity functions return true when an odd number of variables in $S$ are set to $+1$, making them highly non-linear. Majority functions return true if at least half of the input bits are $+1$. Finally, general conjunctions and disjunctions extend their monotone counterparts by allowing negated literals, increasing their expressive capacity and representational complexity.

We tested these function classes across five input sizes: $n = 5, 10, 20, 30, 50$. For each configuration, we ran multiple independent trials—30 for the standard regime and 5 for each constrained or distributional variant—to account for stochastic variability. Hypotheses were encoded as binary strings, where each bit indicates the inclusion or exclusion of a particular literal, and in the case of general functions, a separate bit encodes whether the literal is negated. For example, a general conjunction might be represented as $x_1 \land \lnot x_3 \land x_5$ using a vector like $(1, 0, 1)$ and a polarity vector like $(0, 1, 0)$. 

At each generation, we sampled $s = 1000$ inputs from a distribution $D_n$ over $\{-1, 1\}^n$ and computed the empirical performance of each candidate hypothesis with respect to the target function. The tolerance $t = 0.01$ was chosen to ensure that only statistically meaningful improvements in performance were considered beneficial. Additionally, we used a validation set of 5000 inputs to evaluate whether a hypothesis had successfully evolved. The error parameter was set to $\epsilon = 0.05$, and a hypothesis was deemed successful if it achieved performance greater than $1 - \epsilon = 0.95$ on the validation set.

We imposed a maximum generation cap of 500 iterations. This choice was informed by empirical observations that evolvable function classes typically converged well before this threshold, while unevolvable classes like parity would plateau far earlier. A function class was considered evolvable at a given problem size if at least 3 out of the 5 trials reached the performance threshold on the validation set.

\subsection*{Evolvability Simulation Algorithm}

Below is the general evolvability loop used across all experiments:

\begin{algorithm}[H]
\caption{Evolvability Simulation}
\begin{algorithmic}[1]
\State Initialize target function $f \sim \mathcal{F}_n$ from selected class
\State Sample initial hypothesis $r_0 \in R_n$
\For{$t = 1$ to $T$ (max 1000 generations)}
    \State Generate neighborhood $N(r_{t-1}, \epsilon)$ of size $\leq p(n, 1/\epsilon)$
    \State Sample $s = 1000$ examples $\{x_i\} \sim D_n$
    \State Compute $\text{Perf}_f(r, D_n, s)$ for each $r \in N(r_{t-1}, \epsilon)$
    \State Compute beneficial and neutral sets using $t = 0.01$
    \If{beneficial candidates exist}
        \State Choose $r_t$ from beneficial candidates
    \ElsIf{neutral candidates exist}
        \State Choose $r_t$ from neutral candidates
    \Else
        \State Terminate (no valid mutations)
    \EndIf
    \State Evaluate $r_t$ on 5000 validation points
    \If{$\text{Perf}_f(r_t, D_n) > 0.95$}
        \State \Return Success
    \EndIf
\EndFor
\State \Return Failure
\end{algorithmic}
\end{algorithm}

\subsection*{Distribution Sampling}

In this part of our project, we investigated whether function classes that are evolvable under the uniform distribution retain this property under alternative input distributions. This exploration was motivated by Valiant’s observation that distribution-independent evolvability is particularly desirable, as it ensures robustness to environmental variation and permits continued adaptation even as input patterns shift.

To conduct this analysis, we modified only the distribution from which examples were drawn in each generation of the evolutionary process. All other parameters and mechanisms—such as the mutation strategy, neighborhood generation, tolerance threshold, sample size, and success criteria—remained unchanged. Specifically, we replaced the standard uniform distribution with three distinct alternatives: a binomial distribution, a beta distribution, and a biased Bernoulli distribution with parameter $p = 0.75$. In the binomial case, inputs were sampled from a discrete binomial distribution with parameters chosen to create low-entropy binary vectors. For the beta distribution, continuous samples were drawn from a $\text{Beta}(2,5)$ distribution and then discretized and rescaled into the $\{-1, +1\}^n$ Boolean domain. In the biased Bernoulli case, each bit $x_i$ in the input vector was drawn independently from a $\text{Bernoulli}(0.75)$ distribution and then mapped to $+1$ or $-1$ accordingly. These three configurations allowed us to test whether evolvability persists when the input space is skewed, imbalanced, or clustered, as might occur in natural or real-world settings. The rest of the simulation—including the fitness function, the selection rule, the empirical performance computation, and the success threshold—was left intact, allowing for a clean comparison that isolates the effect of distributional shift.

\subsection*{Constrained Models}

Next, we explored how restricting aspects of the evolvability framework affects the likelihood of successful evolution. This step was motivated by questions Valiant posed in the conclusion of his paper, in which he suggested that relaxing the requirement of arbitrary initial hypotheses or disallowing neutral mutations could significantly alter the dynamics of evolution. We focused on two variants: disallowing neutral mutations and initializing from a fixed hypothesis rather than a random one.

To study the effect of disallowing neutral mutations, we modified the mutator logic so that only beneficial mutations—those that strictly improved empirical performance by at least $t = 0.01$—were accepted. If no such mutation existed in the neighborhood, the evolutionary process terminated early. This change eliminated the fallback to neutral candidates that is permitted in the original framework, resulting in a greedier form of evolution that is potentially more brittle in flat fitness landscapes. The mutation operator, neighborhood size, and all other elements of the simulation remained unchanged from previous parts, ensuring that this experiment isolated the impact of forbidding neutrality.

For the fixed-initialization variant, we replaced the randomly sampled initial hypothesis $r_0$ with a pre-specified one. In the case of conjunctions and disjunctions—both monotone and general—we used the empty hypothesis, which includes no variables and therefore returns false on all inputs. For the majority class, we defined a fixed subset of 10 variables and initialized the hypothesis to compute the majority over just this subset. This ensured a structurally simple and suboptimal starting point that was still meaningful, allowing us to test whether relaxing the requirement of arbitrary initialization reduces the evolvability of a class. Aside from the initialization, the evolutionary loop and evaluation criteria remained identical to those previously.

Across both constrained model variants, we preserved the same sampling distributions, mutation mechanisms, and success thresholds as before. This design allowed us to cleanly assess the impact of removing neutral drift and restricting the starting hypothesis on the overall convergence and evolvability of each function class.

\subsection*{Neighborhood Design for Majority}

A notable implementation challenge arose with the majority function class, as its canonical representation—a threshold over all $n$ variables—is difficult to mutate meaningfully in a localized, polynomial-bounded way. To address this, we defined a neighborhood for majority based on a selected set of “relevant” variables. Each candidate hypothesis was defined over a fixed-size subset of $k$ variables (typically $k = 10$), and mutations involved toggling individual variables in or out of this subset. This approach preserved the structure of majority-based decision-making while allowing for gradual evolutionary improvement in a manner compatible with the theoretical constraints of Valiant’s model.

\subsection*{Code Architecture and Logging}

Our implementation followed a modular design aligned with Valiant’s formal framework. The file \texttt{\path{functions.py}} defines each Boolean function class and generates target functions based on randomly sampled support sets. The file \texttt{\path{environment.py}} provides routines for sampling inputs from $D_n$ and evaluating empirical performance. The core mutation and selection logic is implemented in \texttt{\path{evolve.py}}, where we construct neighborhoods, compute fitness, and select candidates per the mutator definition. Experiments are orchestrated using \texttt{\path{run_experiments.py}}, which executes all trials, records convergence statistics, and writes logs for later analysis. Finally, \texttt{\path{visualization.py}} plots performance curves, generation counts, and evolvability status across all configurations.

\subsection*{Evaluation Criteria}

To evaluate convergence, we monitored the empirical performance of the current hypothesis in each generation. A trial was considered successful if the final performance exceeded $0.95$ on a validation set of 5000 examples within 1000 generations. A function class was considered evolvable under a specific configuration if three or more out of five trials met this success criterion. This threshold balances the need for consistency with tolerance for randomness, ensuring that our conclusions are robust without requiring perfection across all runs.

By articulating these design choices clearly and grounding them in both theory and implementation, we present a rigorous and extensible framework for empirically testing evolvability across a range of function classes, distributions, and modeling assumptions.

\section{Results}

We evaluated six Boolean‐function classes—General Conjunction, General Disjunction, Monotone Conjunction, Monotone Disjunction, Majority, and Parity—under four experimental regimes: standard uniform sampling with neutral mutations allowed; smart initialization; strictly beneficial mutations only; and varied input distributions (Binomial, Beta, Bernoulli(0.75)).  For each function class and each input size $n\in\{5,10,20,30,50\}$, we performed 30 independent runs using tolerance $\epsilon=0.05$, up to 500 generations per run, and 1,000 samples per generation.  We tracked the fraction of runs reaching the target accuracy (success rate), the average number of generations to convergence, the average counts of beneficial versus neutral mutations per generation, and the full fitness‐over‐generations trajectory.

\subsection{Standard (Uniform) Experiments}

\begin{figure}[ht]
  \centering
  \begin{subfigure}[b]{0.45\textwidth}
    \includegraphics[width=\textwidth]{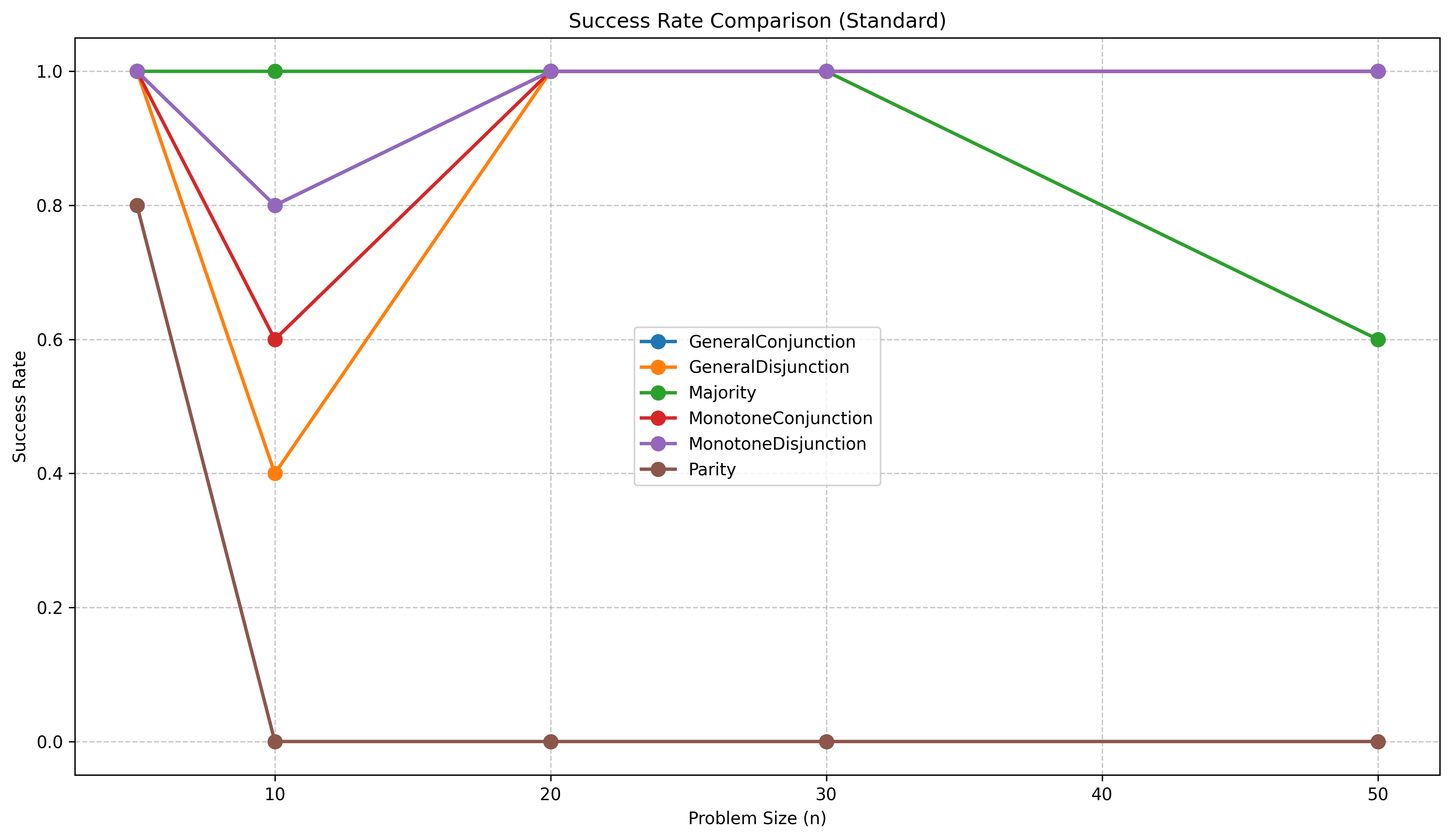}
    \caption{Success Rate vs.\ $n$}
    \label{fig:success_rate}
  \end{subfigure}
  \hfill
  \begin{subfigure}[b]{0.45\textwidth}
    \includegraphics[width=\textwidth]{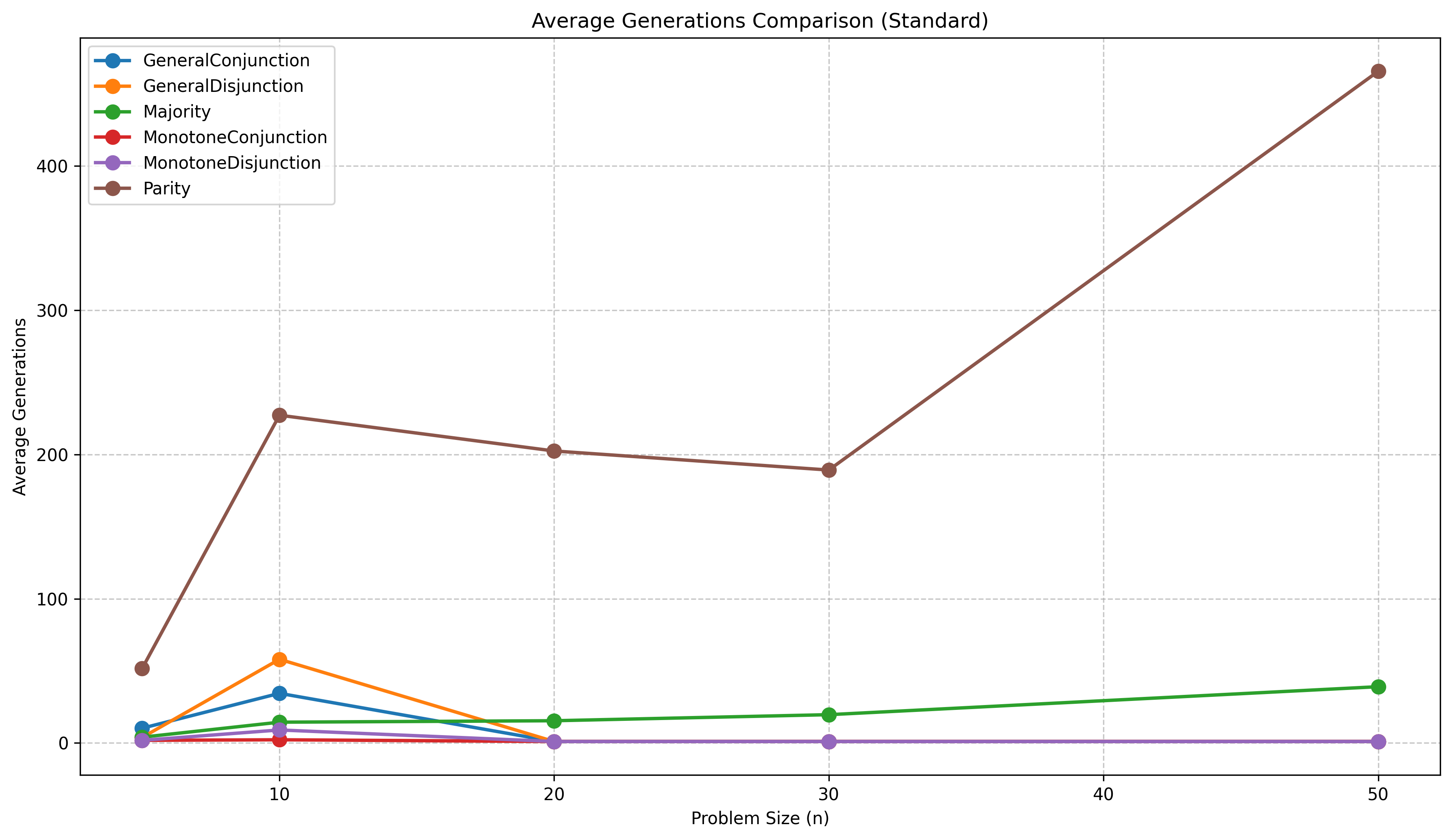}
    \caption{Average Generations}
    \label{fig:avg_gens}
  \end{subfigure}

  \vspace{1em}

  \begin{subfigure}[b]{0.45\textwidth}
    \includegraphics[width=\textwidth]{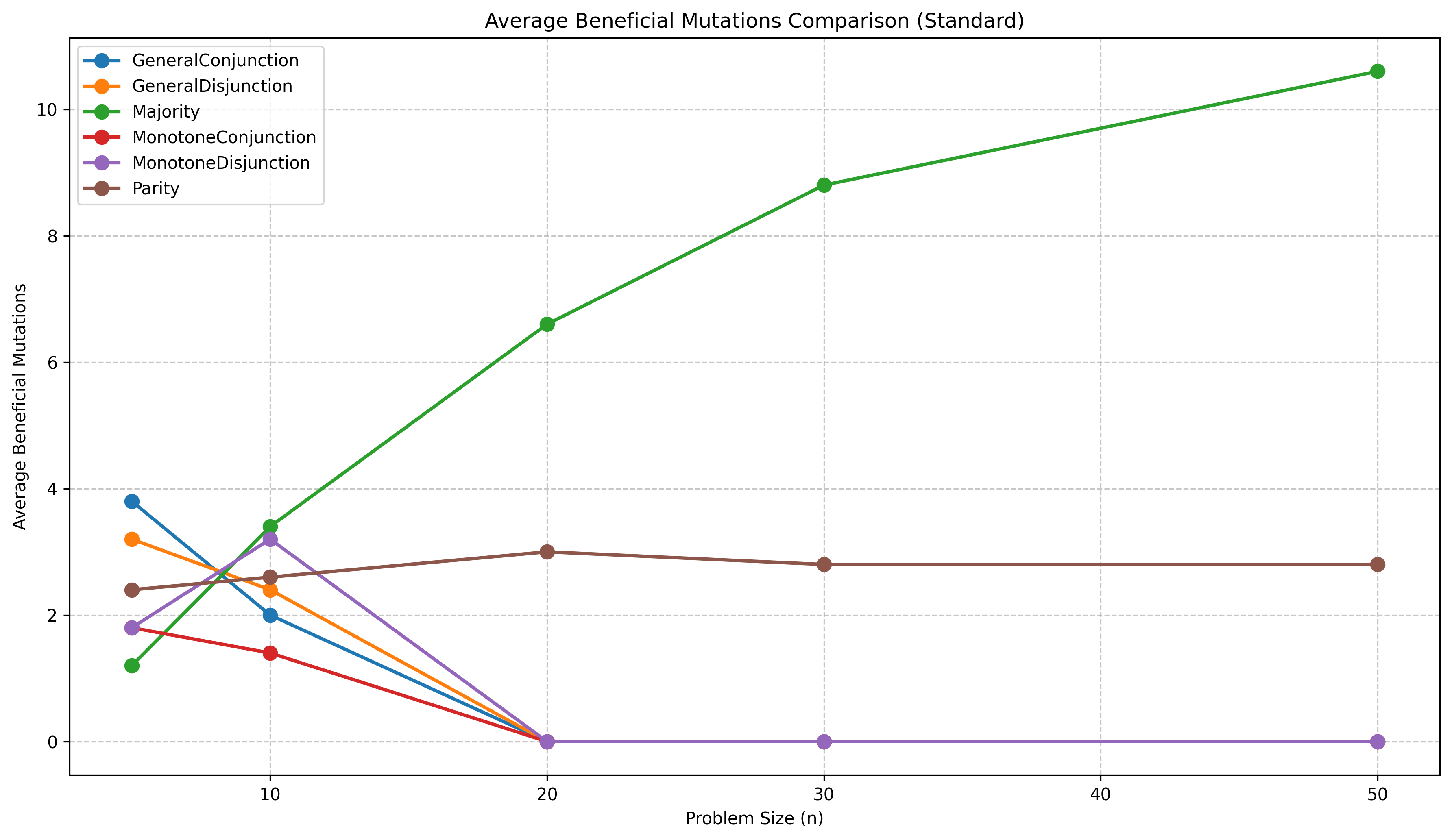}
    \caption{Beneficial Mutations per Generation}
    \label{fig:beneficial}
  \end{subfigure}
  \hfill
  \begin{subfigure}[b]{0.45\textwidth}
    \includegraphics[width=\textwidth]{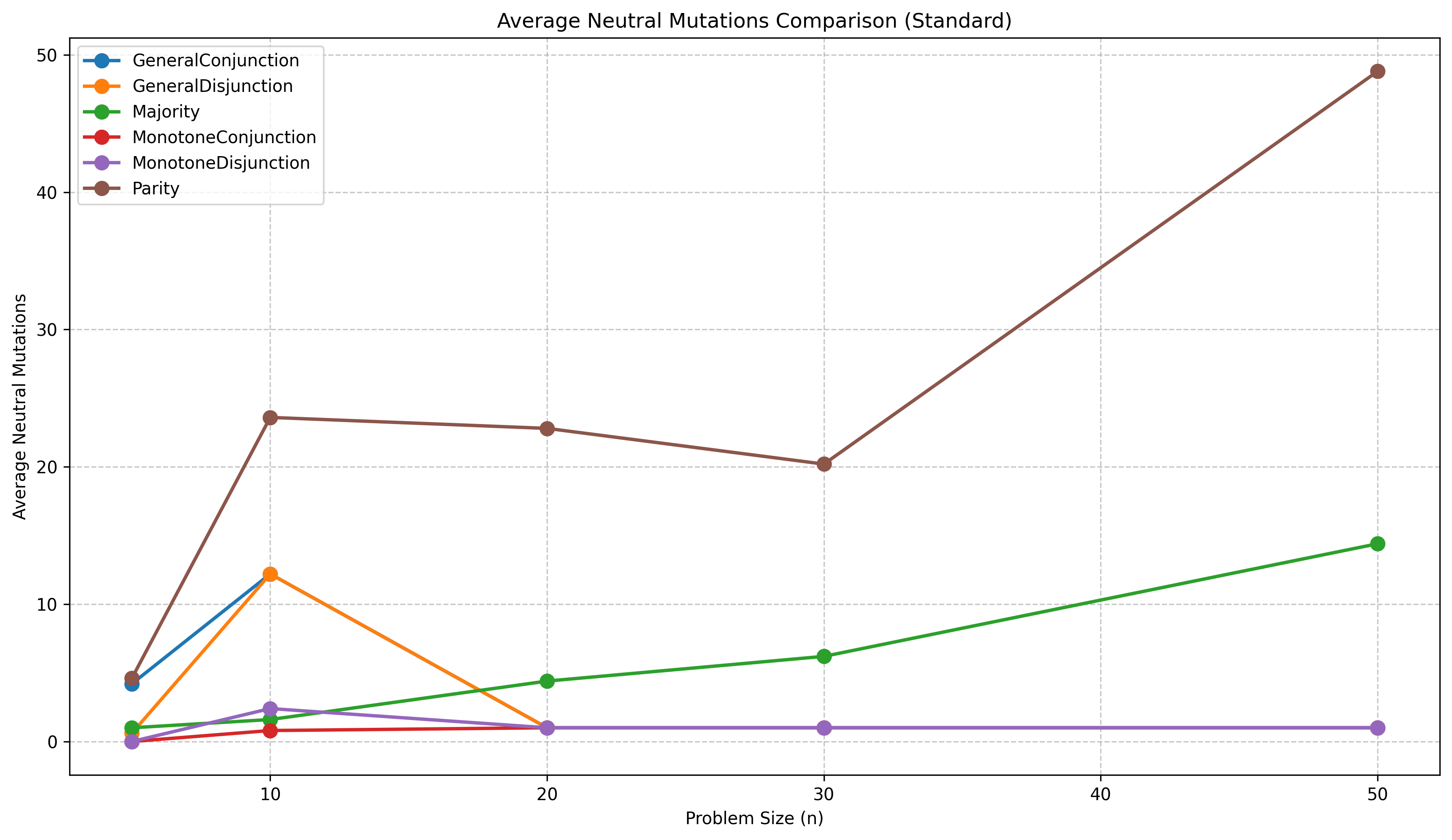}
    \caption{Neutral Mutations per Generation}
    \label{fig:neutral}
  \end{subfigure}

  \caption{Summary metrics under uniform–Bernoulli sampling.}
  \label{fig:standard_summary}
\end{figure}

All input bits were drawn i.i.d.\ from Bernoulli(0.5).  Figure~\ref{fig:standard_summary} presents the four summary metrics in a 2×2 layout: the top‐left panel shows success rate as a function of $n$, the top‐right shows average generations until convergence (for successful runs), the bottom‐left shows average beneficial mutations per generation, and the bottom‐right shows average neutral mutations per generation.

\FloatBarrier

\begin{figure}[ht]
  \centering
  \begin{subfigure}[b]{0.45\textwidth}
    \includegraphics[width=\textwidth]{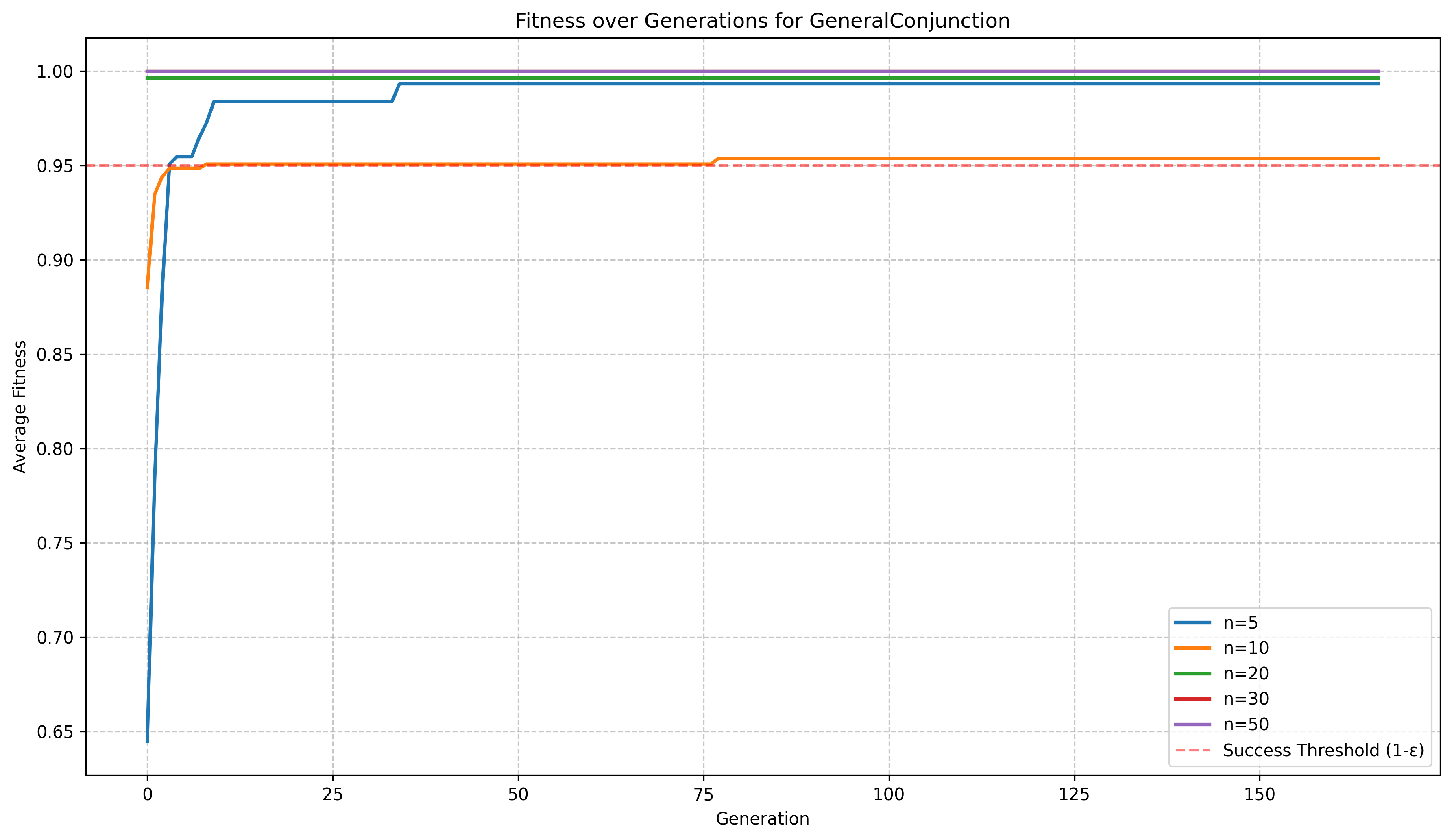}
    \caption{General Conjunction}
  \end{subfigure}
  \hfill
  \begin{subfigure}[b]{0.45\textwidth}
    \includegraphics[width=\textwidth]{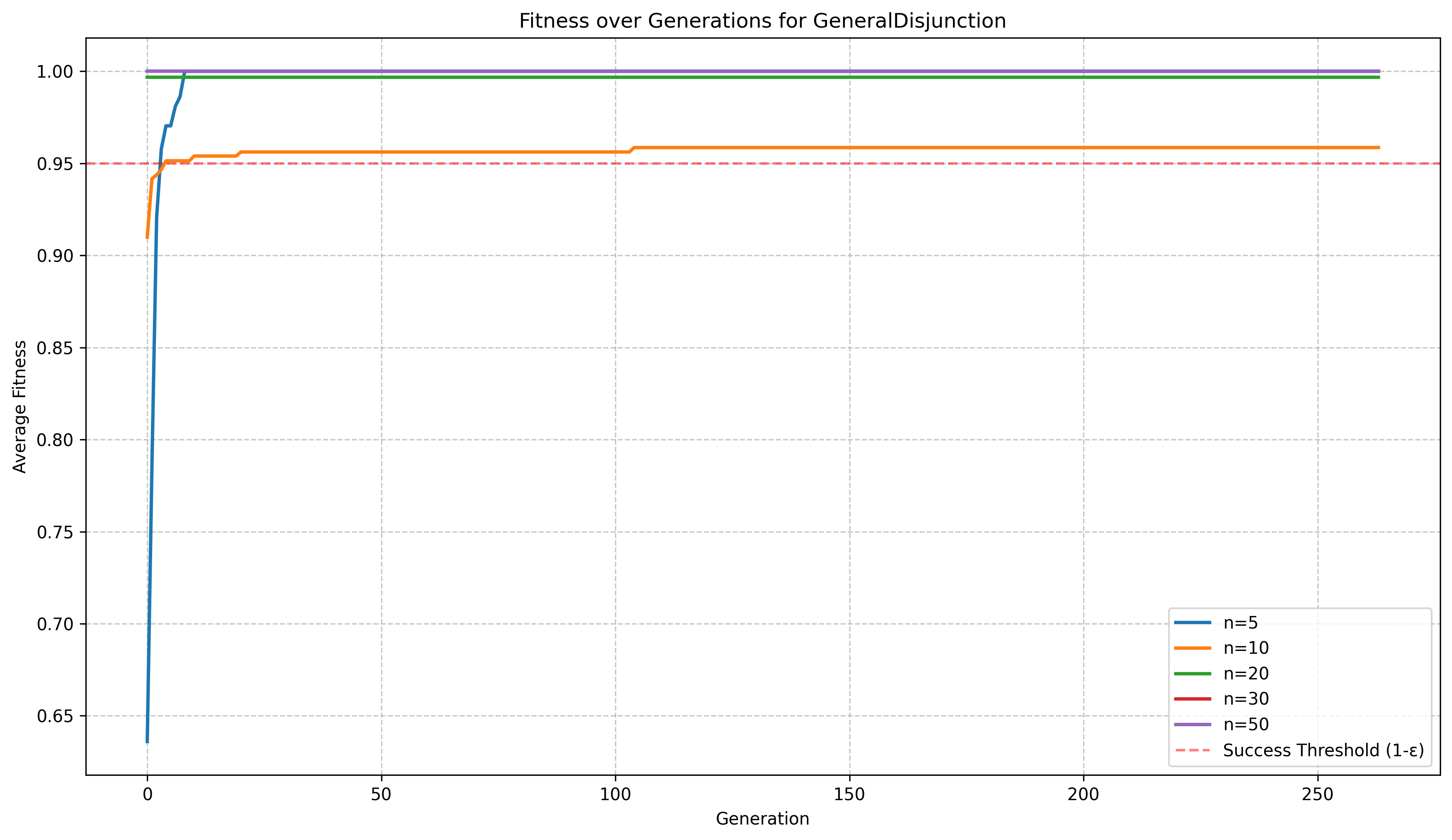}
    \caption{General Disjunction}
  \end{subfigure}

  \vspace{1em}

  \begin{subfigure}[b]{0.45\textwidth}
    \includegraphics[width=\textwidth]{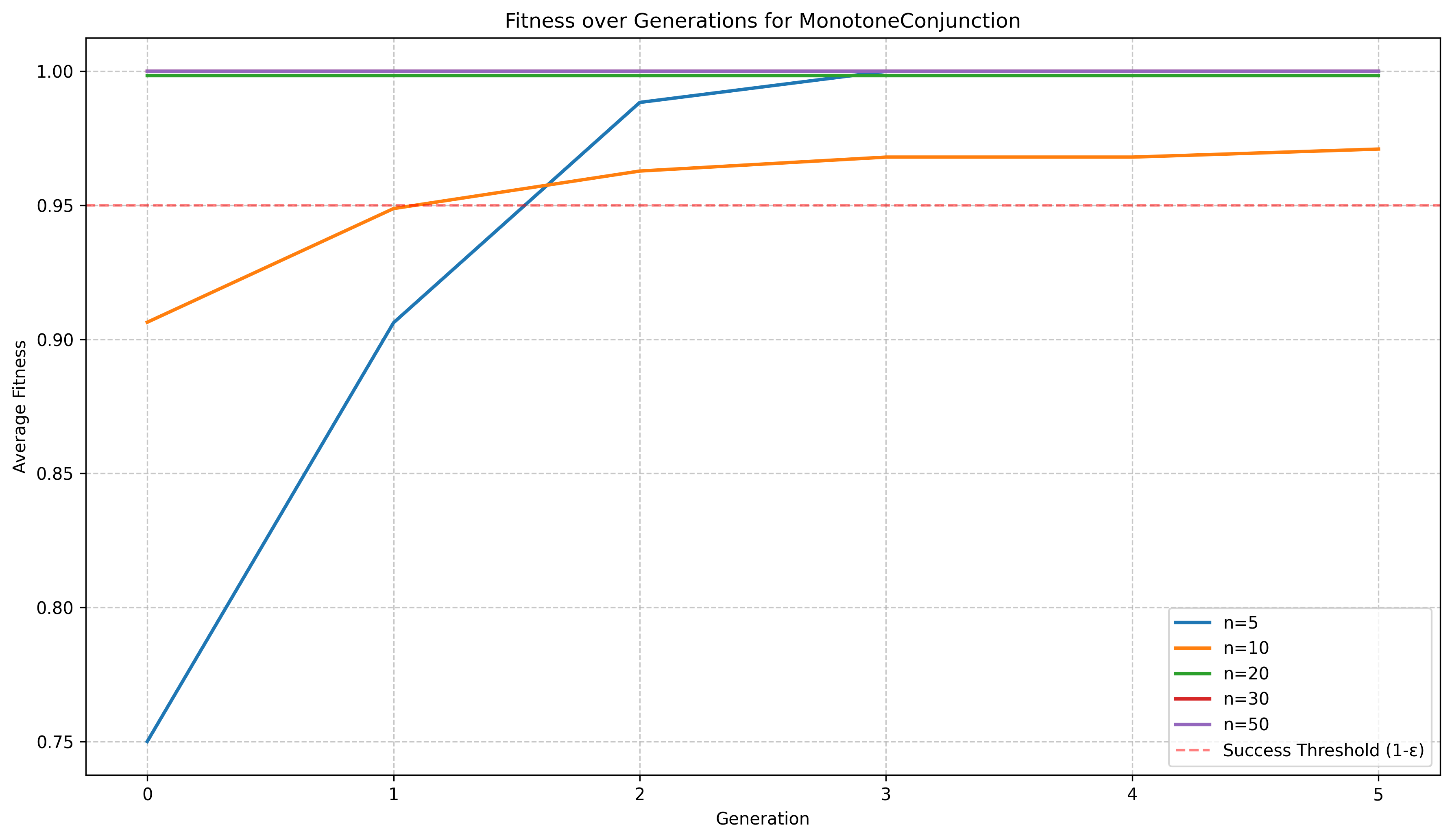}
    \caption{Monotone Conjunction}
  \end{subfigure}
  \hfill
  \begin{subfigure}[b]{0.45\textwidth}
    \includegraphics[width=\textwidth]{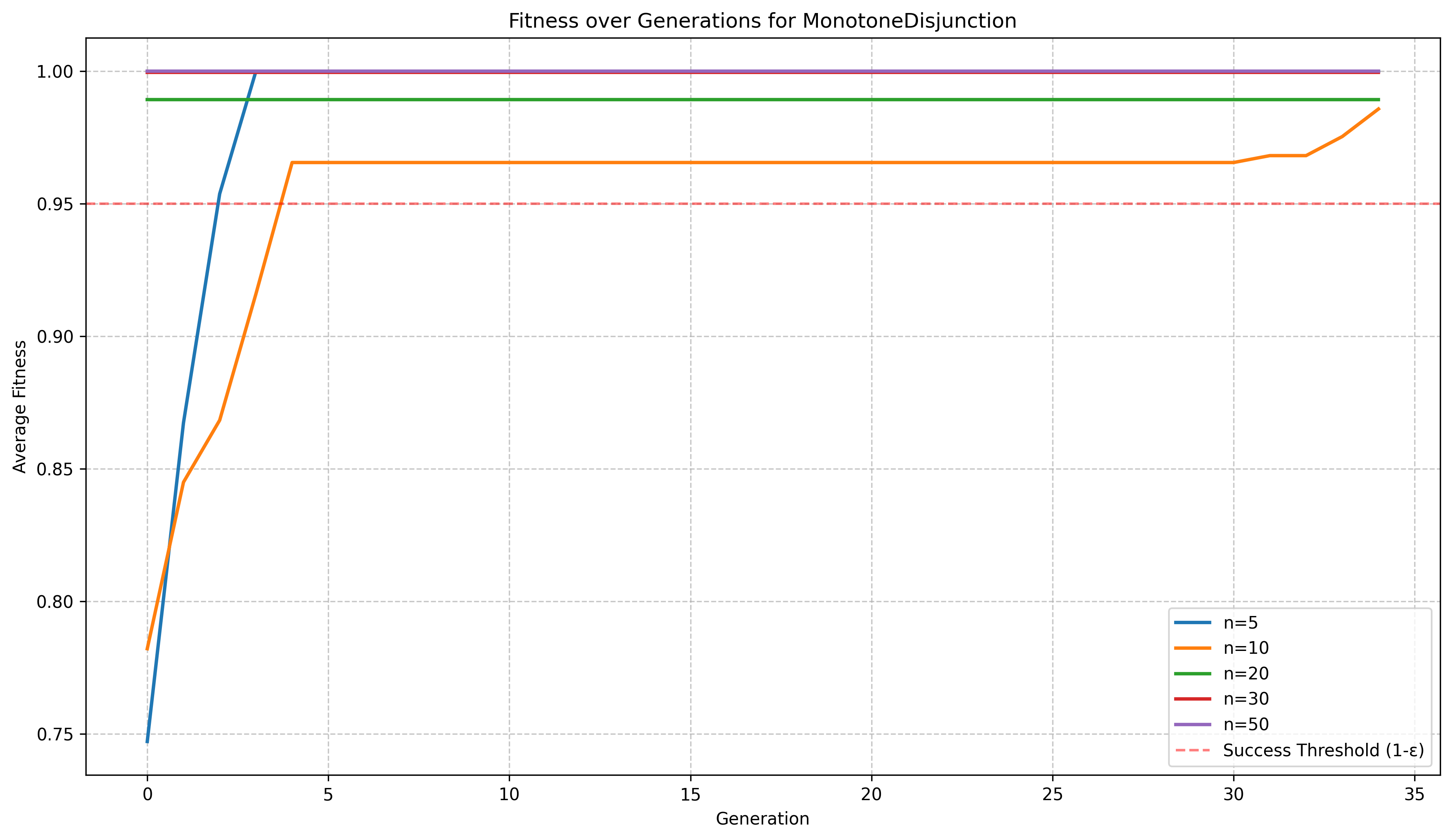}
    \caption{Monotone Disjunction}
  \end{subfigure}

  \vspace{1em}

  \begin{subfigure}[b]{0.45\textwidth}
    \includegraphics[width=\textwidth]{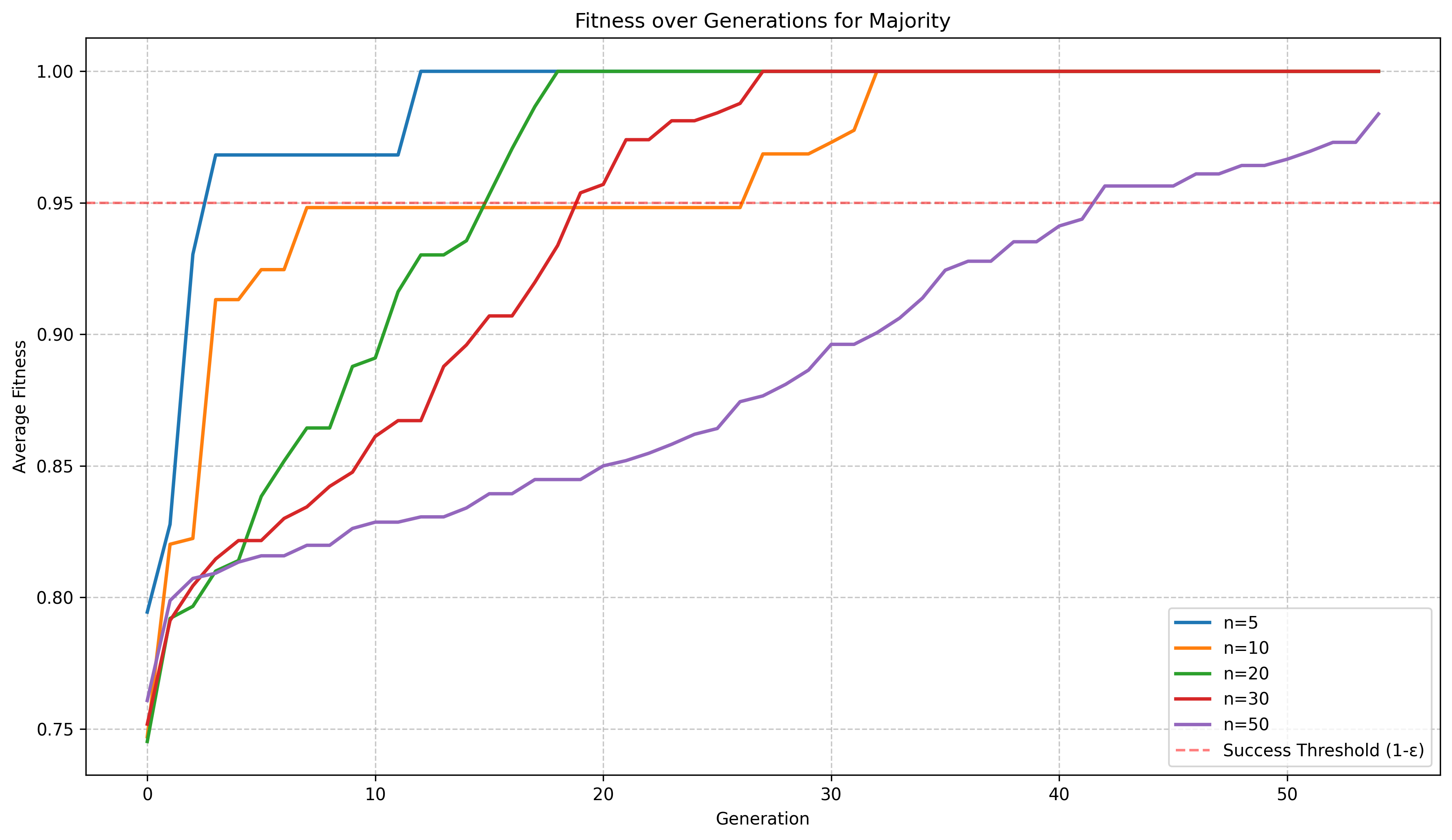}
    \caption{Majority}
  \end{subfigure}
  \hfill
  \begin{subfigure}[b]{0.45\textwidth}
    \includegraphics[width=\textwidth]{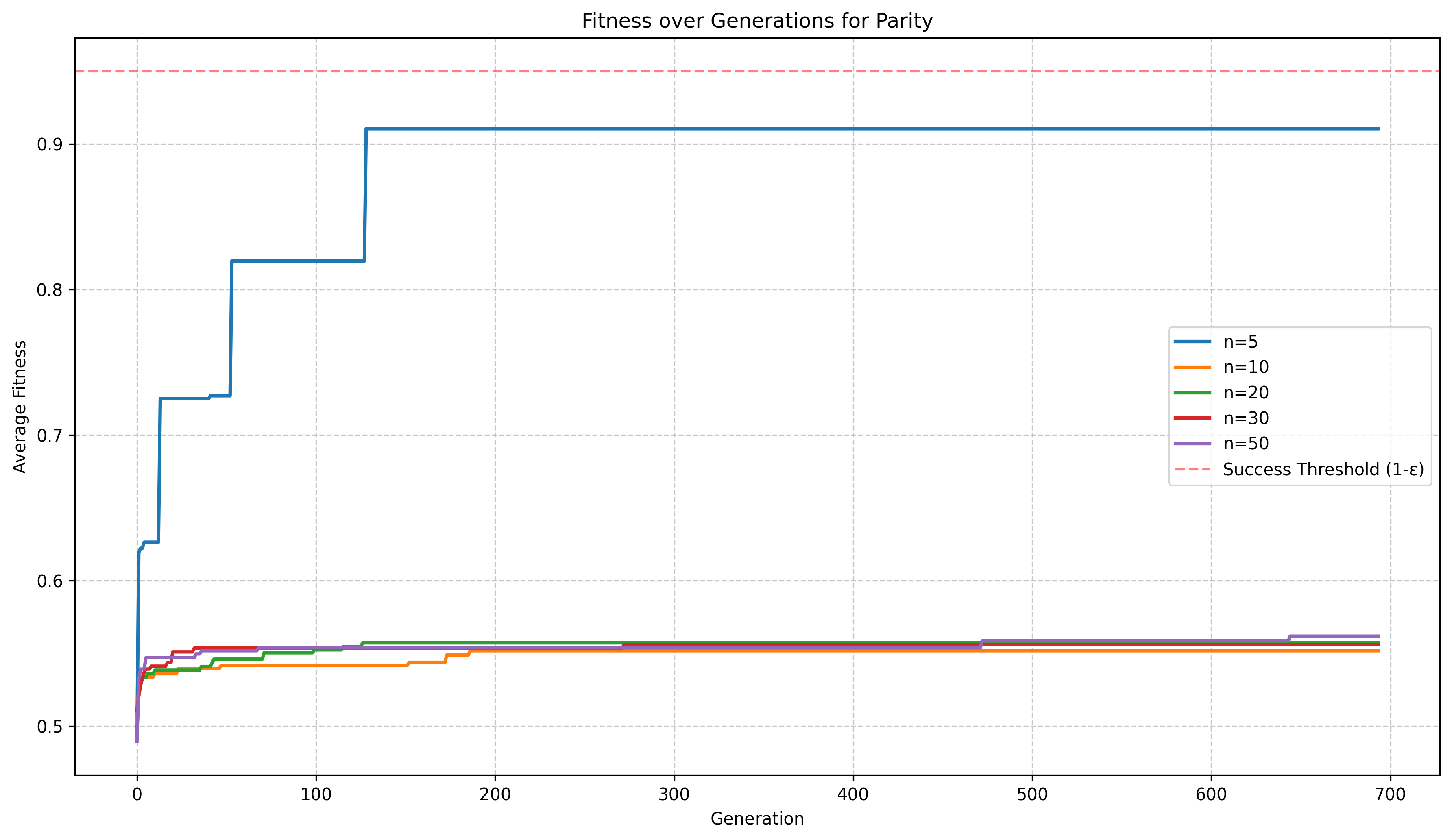}
    \caption{Parity}
  \end{subfigure}

  \caption{Fitness‐over‐generations curves under uniform sampling.}
  \label{fig:standard_fitness}
\end{figure}

Figure~\ref{fig:standard_fitness} shows the mean fitness over generations for each class in a 3×2 grid: General Conjunction and General Disjunction in the first row, Monotone Conjunction and Monotone Disjunction in the second row, and Majority and Parity in the third row.

Monotone Conjunction and Monotone Disjunction reach the target accuracy in approximately 1.0–1.2 generations across all tested dimensions, with nearly vertical fitness trajectories (see Figure~\ref{fig:standard_fitness}c,d).  General Conjunction and General Disjunction converge for most $n$, but both exhibit a pronounced stall at $n=10$, where their fitness curves (Figure~\ref{fig:standard_fitness}a,b) remain near initialization for dozens of generations before rapid ascent.  This results in success rates dropping to 80\% and 40\% and convergence times spiking to ~34 and ~58 generations, respectively.  Majority functions display steady fitness improvements up to $n=30$, then several runs stall around 0.9 accuracy at $n=50$, bringing success down to 60\% (Figure~\ref{fig:standard_fitness}e).  Parity’s fitness curves (Figure~\ref{fig:standard_fitness}f) remain flat, confirming its non‐evolvability beyond $n=5$.

The marked decrease in evolvability at $n=10$ for the general‐class functions coincides with a peak in neutral mutations and a trough in beneficial mutations (Figure~\ref{fig:standard_summary}c,d).  We interpret this as the emergence of large neutral plateaus in the search landscape at this intermediate dimension: the hypothesis‐space size ($2^{10}\approx10^3$) is sufficient to produce many fitness‐invariant neighbors but too small for neutral drift alone to consistently discover the rare beneficial directions.  Moreover, increased noise in fitness estimation at $n=10$ can mask incremental improvements, further extending the plateau periods seen in Figure~\ref{fig:standard_fitness}a,b.

\FloatBarrier

\subsection{Smart Initialization}

\begin{figure}[ht]
  \centering
  \begin{subfigure}[b]{0.45\textwidth}
    \includegraphics[width=\textwidth]{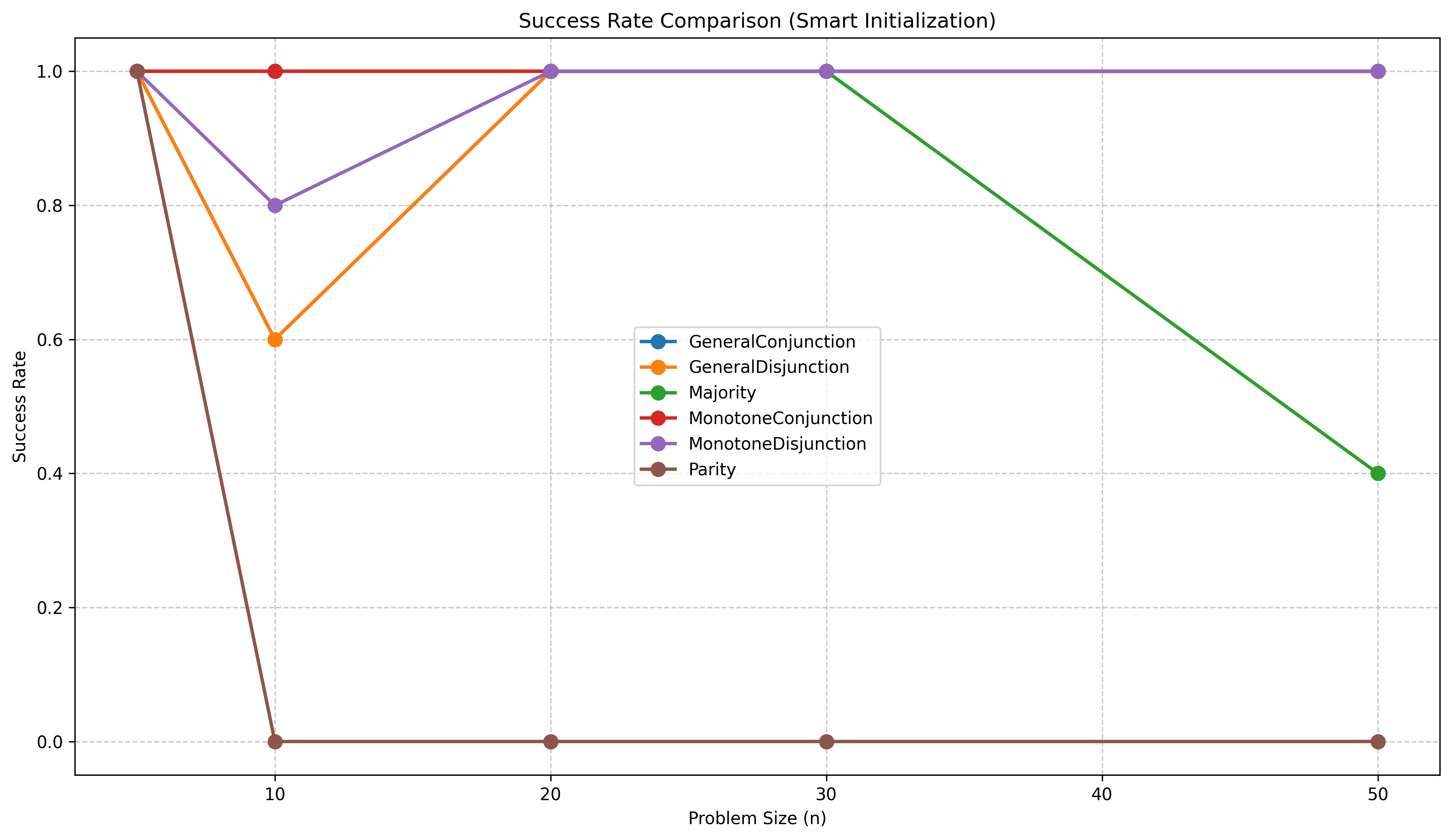}
    \caption{Success Rate vs.\ $n$}
  \end{subfigure}
  \hfill
  \begin{subfigure}[b]{0.45\textwidth}
    \includegraphics[width=\textwidth]{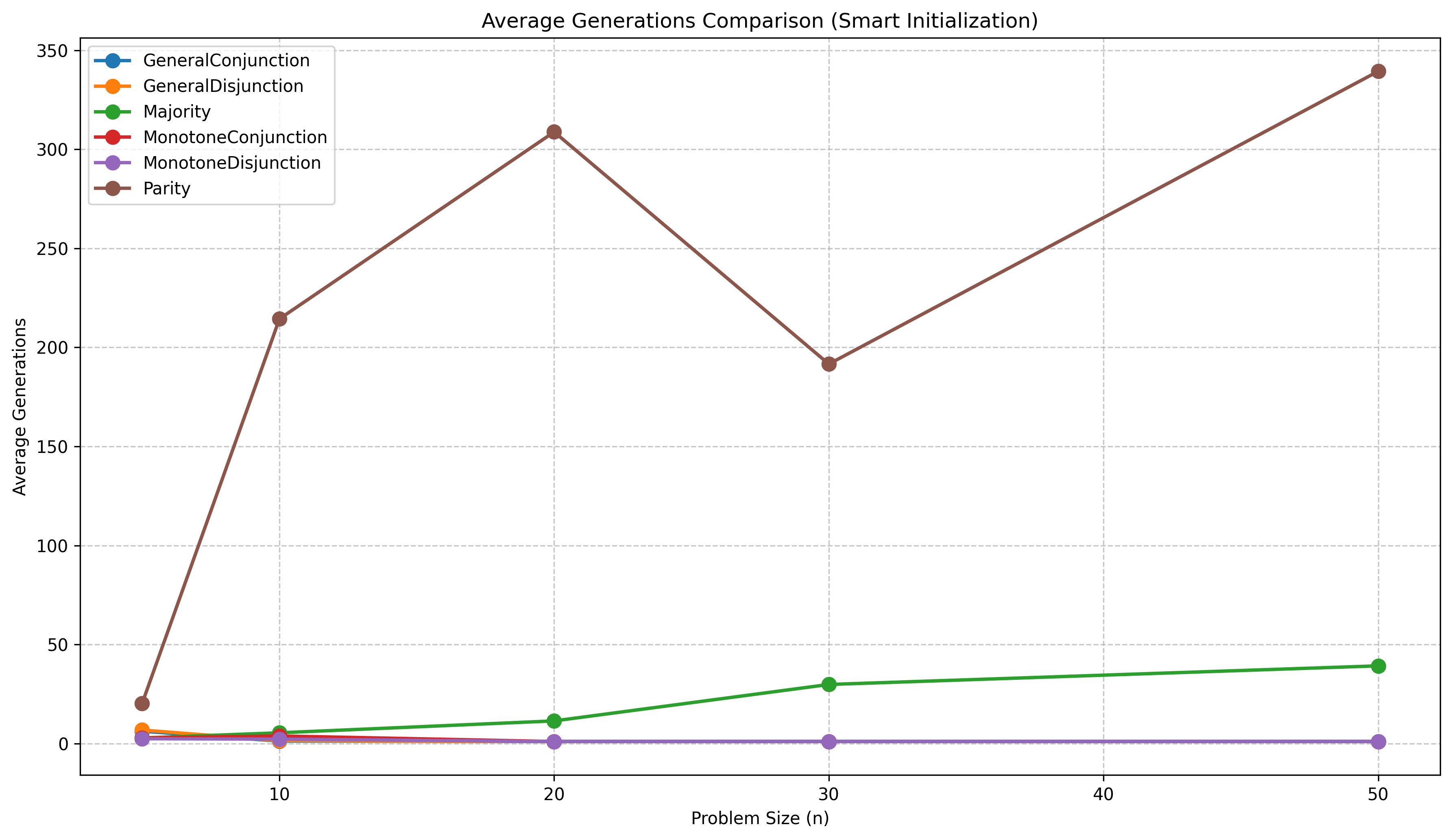}
    \caption{Average Generations}
  \end{subfigure}

  \vspace{1em}

  \begin{subfigure}[b]{0.45\textwidth}
    \includegraphics[width=\textwidth]{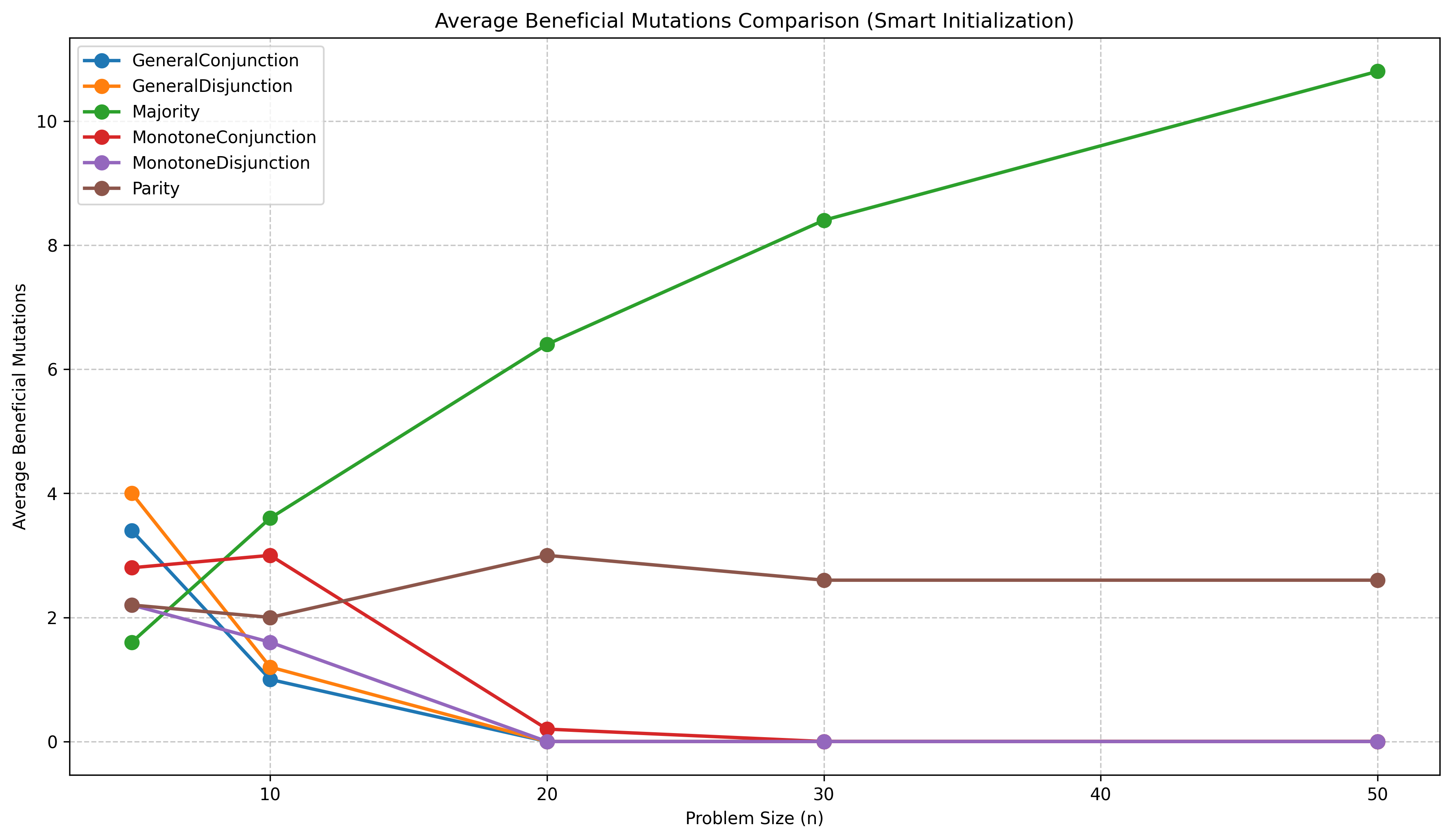}
    \caption{Beneficial Mutations per Generation}
  \end{subfigure}
  \hfill
  \begin{subfigure}[b]{0.45\textwidth}
    \includegraphics[width=\textwidth]{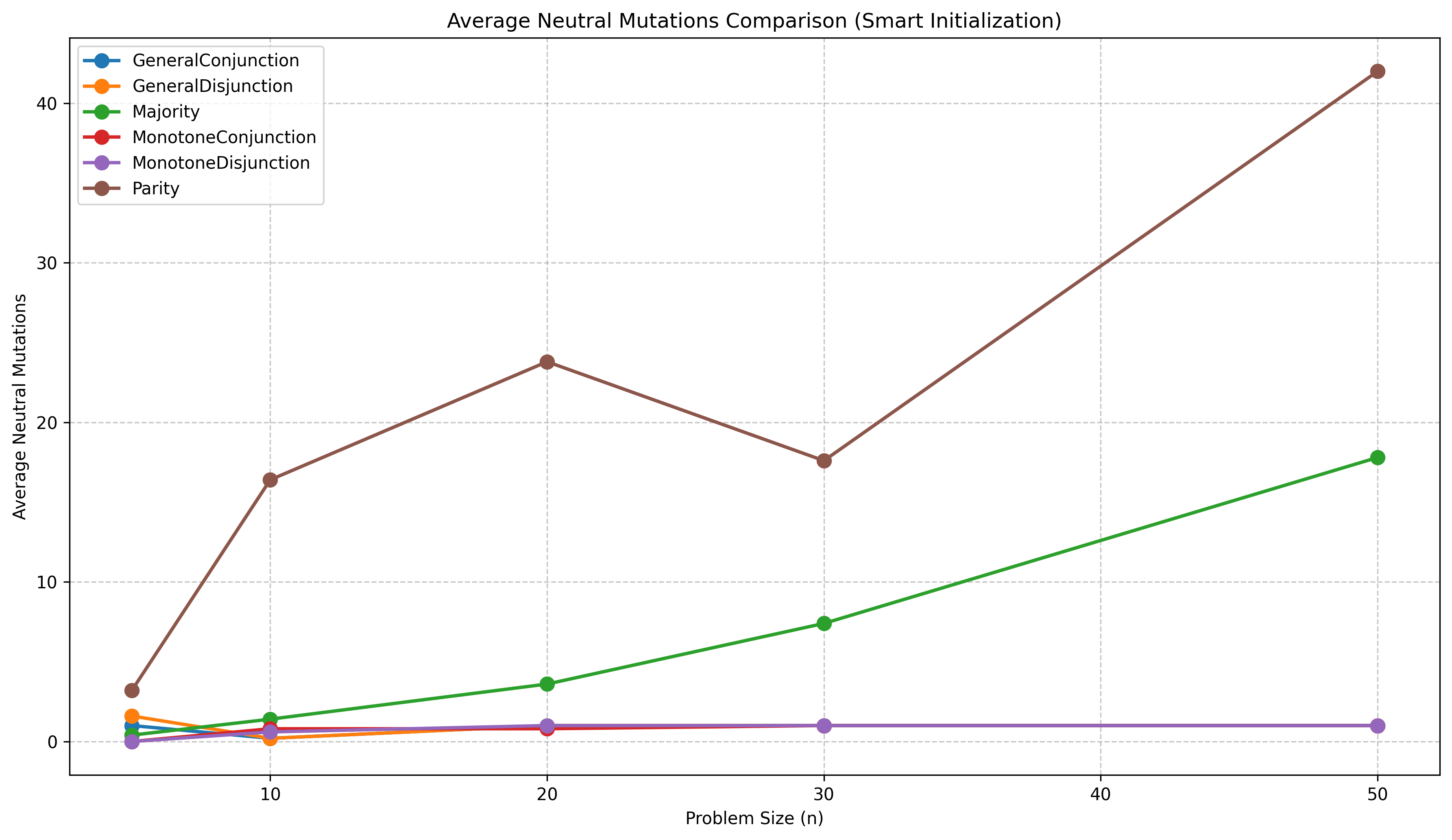}
    \caption{Neutral Mutations per Generation}
  \end{subfigure}

  \caption{Summary metrics under smart initialization.}
  \label{fig:smart_summary}
\end{figure}

Seeding the initial hypothesis to be closer to the target dramatically reduces convergence time but does not change which classes ultimately evolve.  Figure~\ref{fig:smart_summary} arranges the four summary metrics in a 2×2 grid: panel (a) shows success rates versus $n$, panel (b) shows the average number of generations to reach the threshold, panel (c) shows the average count of beneficial mutations per generation, and panel (d) shows the average count of neutral mutations per generation.

Under this regime, General Conjunction and General Disjunction achieve 100\% success for all tested dimensions except $n=10$, where each attains only 60\% success and requires approximately 1.2 and 1.4 generations, respectively, to converge (Figure~\ref{fig:smart_summary}b).  Monotone Conjunction and Monotone Disjunction evolve trivially in about 1.0–1.5 generations with minimal neutral drift (Figure~\ref{fig:smart_summary}c–d).  Majority functions hold 100\% success up to $n=30$ but decline to 40\% at $n=50$, taking on the order of 39 generations when successful.  Parity succeeds only at $n=5$ and fails thereafter, with the few partial runs dominated by neutral moves.

Although smart initialization smooths and accelerates convergence—reducing the average generation counts across all classes—the pronounced dip at $n=10$ for general classes persists.  The elevated neutral‐to‐beneficial mutation ratio at that dimension (Figure~\ref{fig:smart_summary}d) indicates that starting closer to the target does not eliminate the combinatorial plateaus intrinsic to the $n=10$ search space.  

\subsection{No‐Neutral (Strictly Beneficial) Experiments}

\begin{figure}[ht]
  \centering
  \begin{subfigure}[b]{0.45\textwidth}
    \includegraphics[width=\textwidth]{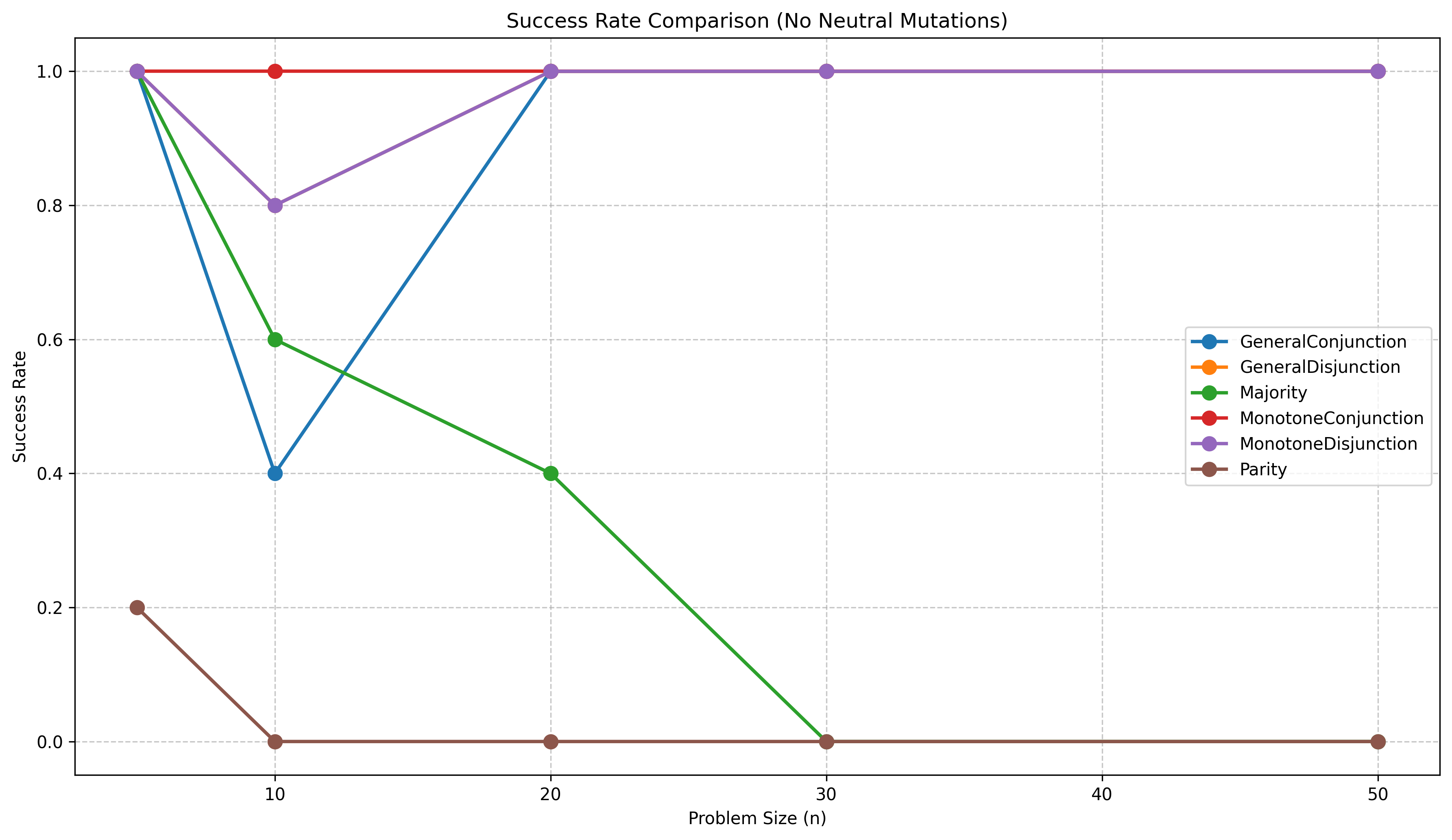}
    \caption{Success Rate vs.\ $n$}
    \label{fig:no_neutral_success}
  \end{subfigure}
  \hfill
  \begin{subfigure}[b]{0.45\textwidth}
    \includegraphics[width=\textwidth]{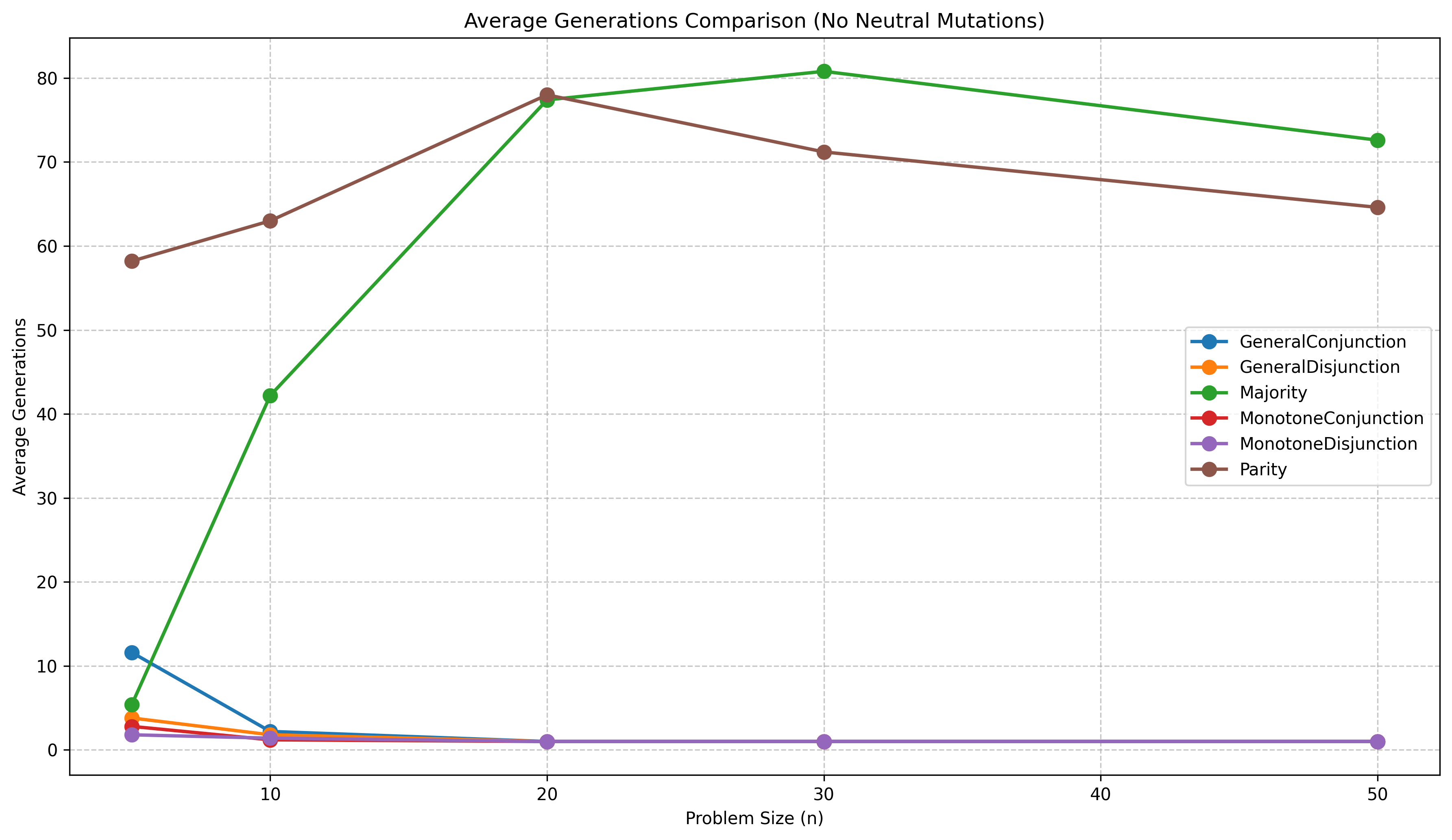}
    \caption{Average Generations}
    \label{fig:no_neutral_gens}
  \end{subfigure}

  \vspace{1em}

  \begin{subfigure}[b]{0.45\textwidth}
    \includegraphics[width=\textwidth]{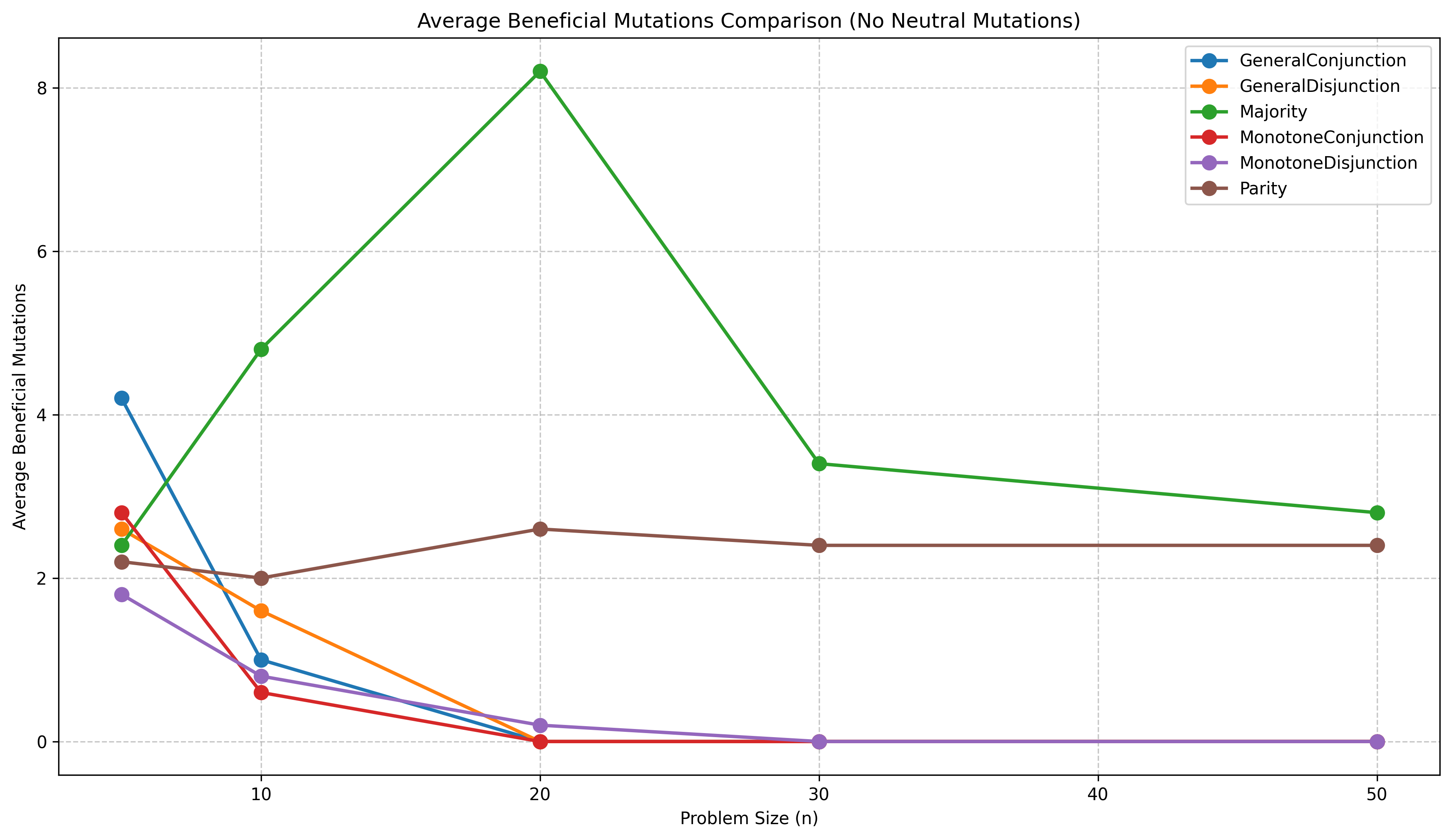}
    \caption{Beneficial Mutations per Generation}
    \label{fig:no_neutral_beneficial}
  \end{subfigure}
  \hfill
  \begin{subfigure}[b]{0.45\textwidth}
    \includegraphics[width=\textwidth]{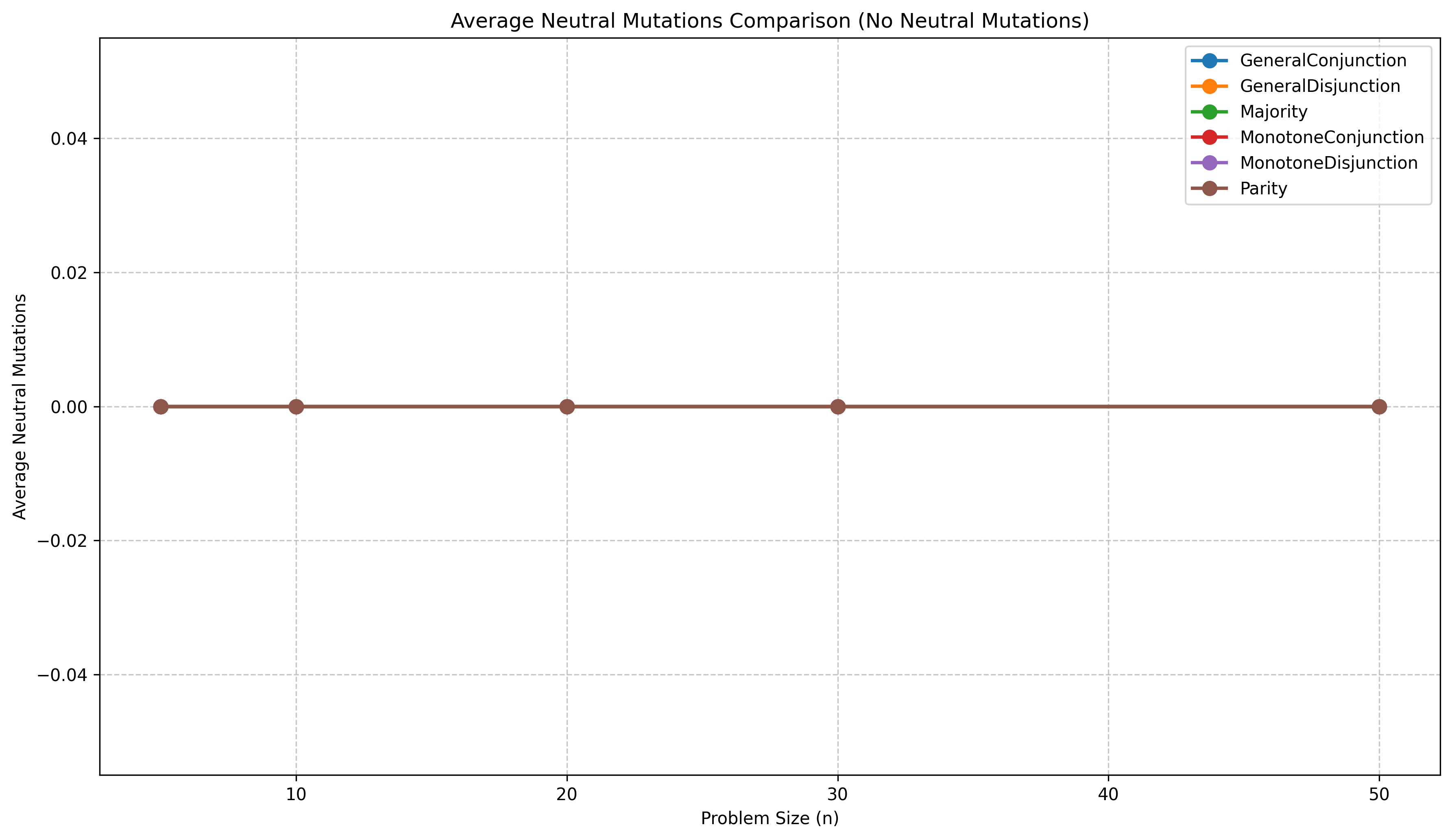}
    \caption{Neutral Mutations per Generation}
    \label{fig:no_neutral_neutral}
  \end{subfigure}

  \caption{Summary metrics under the no‐neutral regime.}
  \label{fig:no_neutral_summary}
\end{figure}

In the no‐neutral regime, only mutations that strictly increased fitness were accepted; all neutral or deleterious mutations were discarded.  Figure~\ref{fig:no_neutral_summary} displays the four summary metrics in a 2×2 panel: success rate, average generations to convergence, beneficial mutations per generation, and neutral mutations per generation.

General Conjunction and General Disjunction continue to evolve for most values of $n$, but both classes suffer marked failures at $n=10$—success rates drop to 40\% and 80\%, respectively—and require more generations to converge than under the neutral‐allowed regime.  As $n$ increases beyond 10, both classes return to near‐perfect success, indicating that larger hypothesis spaces provide enough immediately beneficial moves to compensate for the loss of neutral exploration.  Monotone Conjunction and Monotone Disjunction remain trivial, converging at 100\% success in approximately 1–2 generations regardless of dimension.

Majority functions now fail completely for $n\ge20$, with only 60\% success at $n=10$ and zero successes at larger dimensions.  The few successful runs at $n=10$ consume on the order of 40 generations, but beyond that no beneficial mutation arises frequently enough to escape initial hypotheses.  Parity is entirely intractable under strictly beneficial‐only moves, with no runs reaching the target for any tested dimension.

Removing neutral drift eliminates the exploratory steps needed to traverse flat regions of the fitness landscape.  In the general classes, this deepens the $n=10$ dip compared to the standard regime, since no neutral moves can bridge low‐fitness plateaus to higher‐fitness regions.  For complex classes such as Majority and Parity, these plateaus become insurmountable without tolerance for neutral exploration, resulting in total failure.

\subsection{Distributional Variants}

\begin{figure}[ht]
  \centering
  \begin{subfigure}[b]{0.45\textwidth}
    \includegraphics[width=\textwidth]{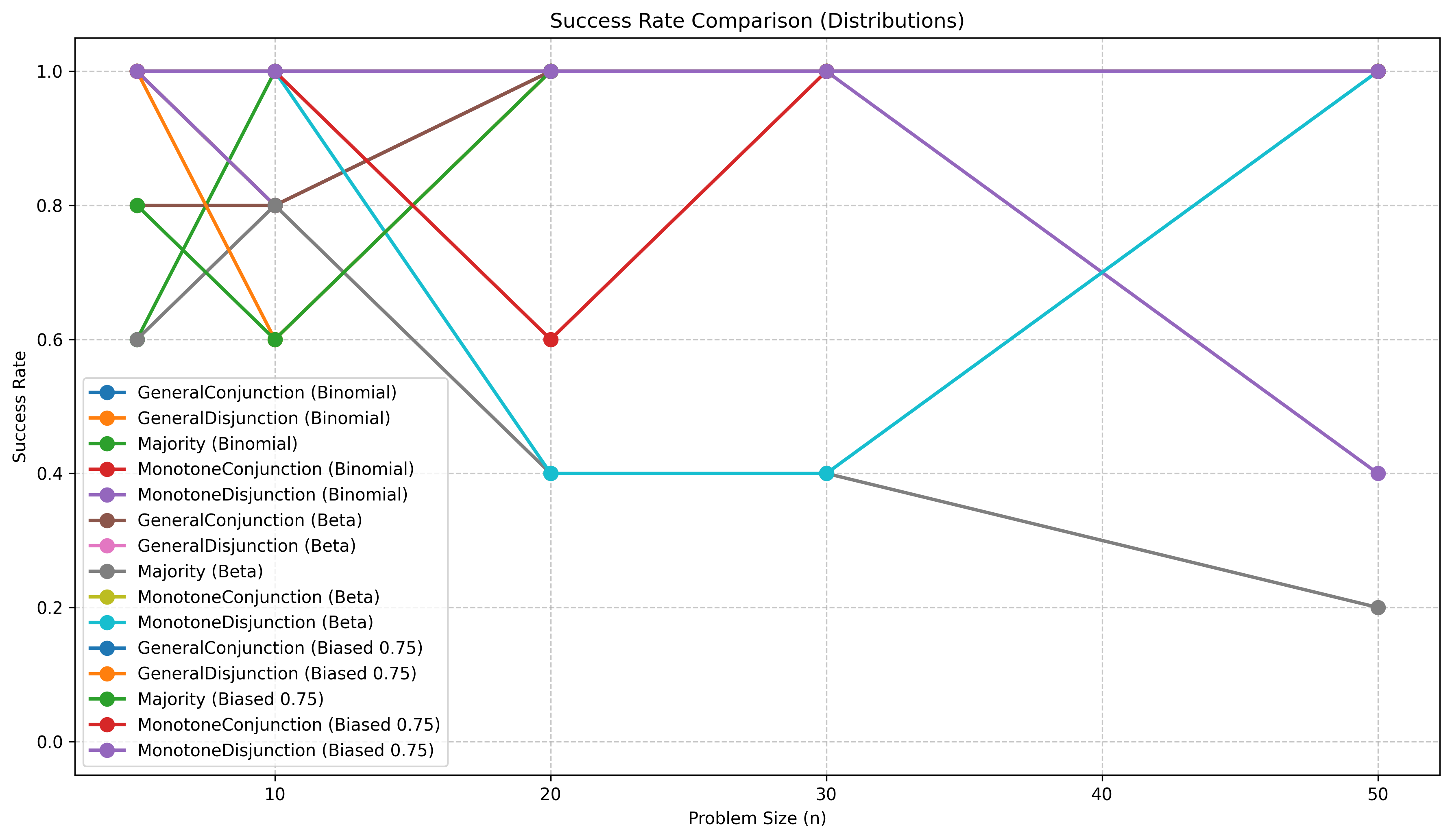}
    \caption{Success Rate}
  \end{subfigure}
  \hfill
  \begin{subfigure}[b]{0.45\textwidth}
    \includegraphics[width=\textwidth]{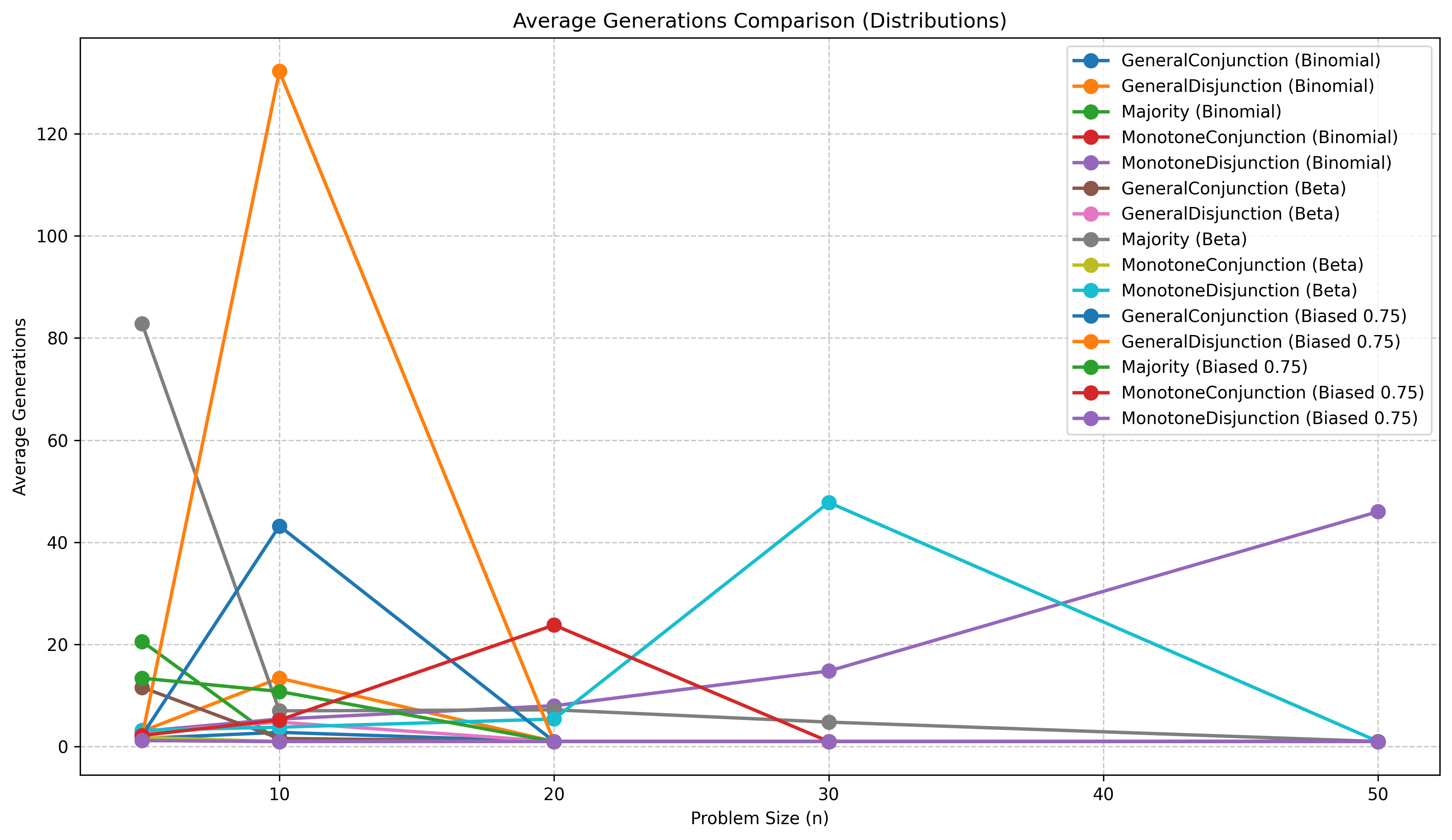}
    \caption{Avg.\ Generations}
  \end{subfigure}

  \vspace{1em}

  \begin{subfigure}[b]{0.45\textwidth}
    \includegraphics[width=\textwidth]{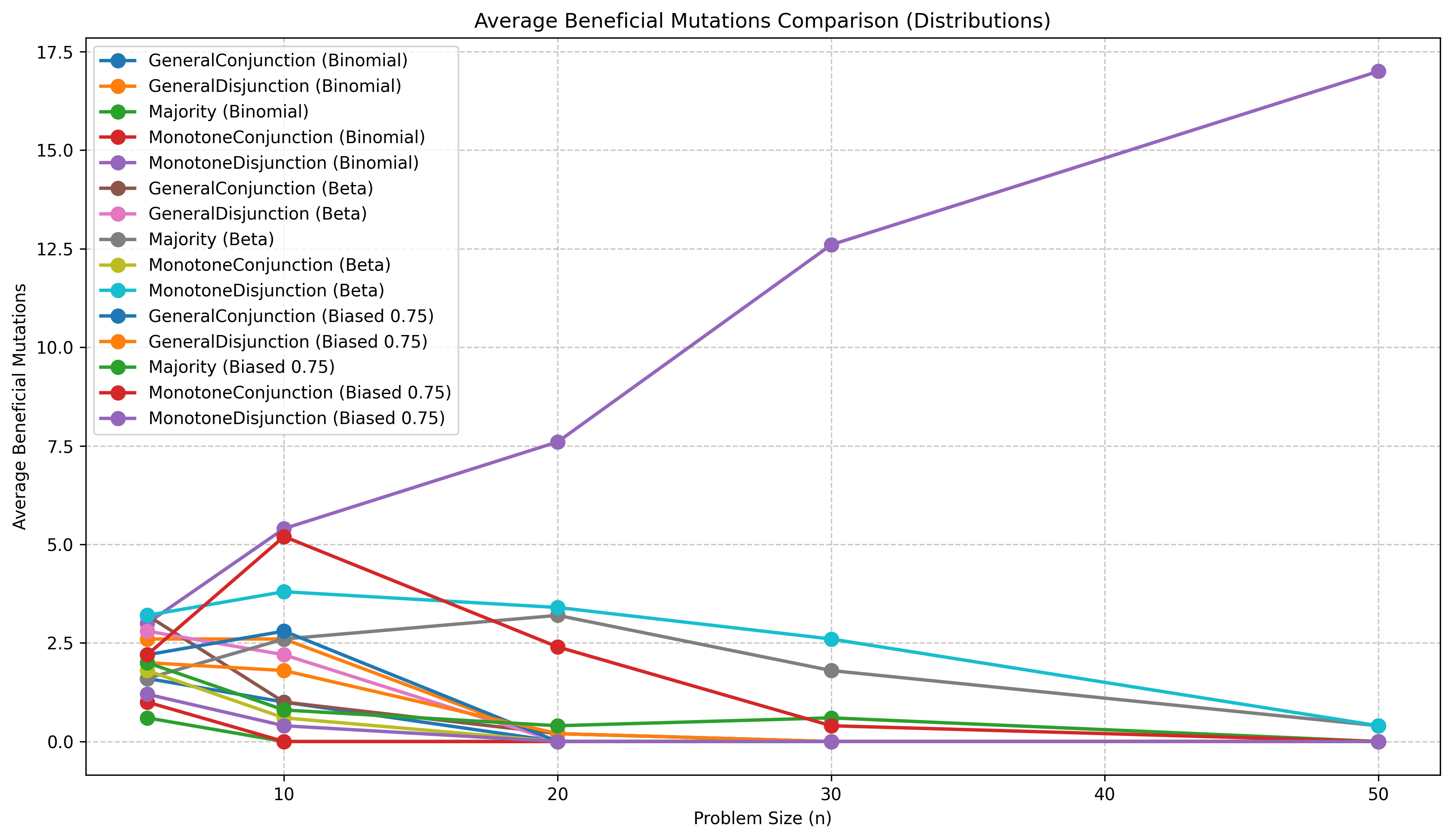}
    \caption{Beneficial Mutations}
  \end{subfigure}
  \hfill
  \begin{subfigure}[b]{0.45\textwidth}
    \includegraphics[width=\textwidth]{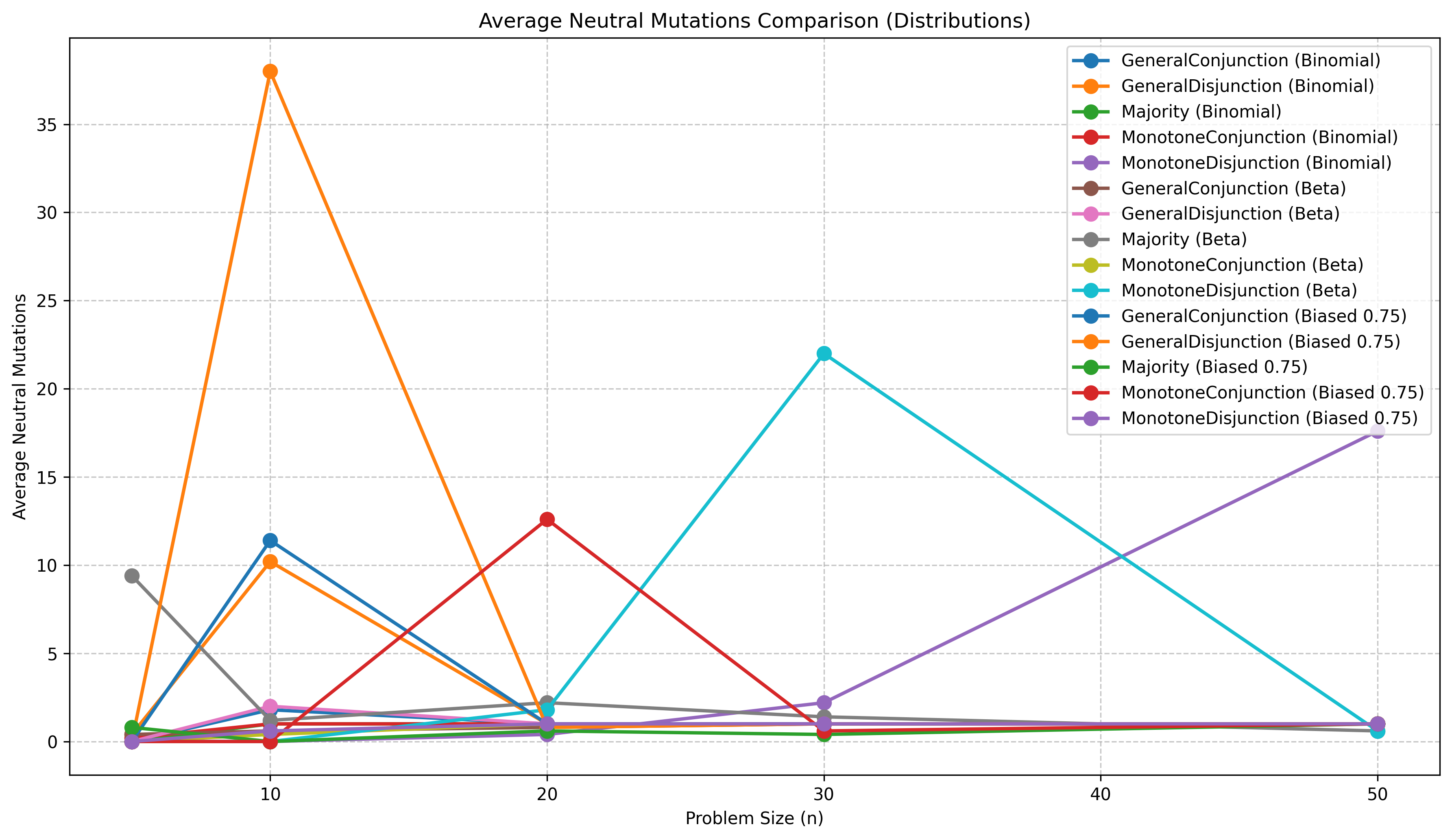}
    \caption{Neutral Mutations}
  \end{subfigure}

  \caption{Summary metrics under alternative sampling regimes: Uniform (baseline), Binomial, Beta, and Bernoulli(0.75).}
  \label{fig:dist_summary}
\end{figure}

We evaluated evolvability under three non‐uniform sampling regimes—Binomial($n$,0.5), Beta(2,5)-driven Bernoulli, and Bernoulli(0.75)—in addition to the uniform baseline.  Figure~\ref{fig:dist_summary} shows the aggregated performance across these distributions in a 2×2 grid: success rate (a), average generations to convergence (b), beneficial mutations per generation (c), and neutral mutations per generation (d).

General Conjunction and General Disjunction consistently achieve 100\% success across all four sampling regimes, and Monotone Conjunction/Disjunction likewise converge in a single generation in each case.  Majority functions are sensitive to input skew: success remains at 100\% under the Binomial and Biased(0.75) regimes but drops to 20\% under the Beta‐skewed regime.  Parity is non‐evolvable under every distribution, averaging hundreds of generations without reaching the accuracy threshold.

Figure~\ref{fig:dist_summary}(b) shows that average generations for the trivial classes remain at approximately one, and for Majority under Binomial and Biased sampling, while the few successful Beta‐runs also converge quickly ($\approx$1 generation).  Figure~\ref{fig:dist_summary}(c,d) reveals that under Beta sampling, Majority exhibits a high neutral‐to‐beneficial mutation ratio, indicating a rugged, plateau‐filled fitness landscape when inputs are sparse.

\begin{figure}[ht]
  \centering
  \begin{subfigure}[b]{0.45\textwidth}
    \includegraphics[width=\textwidth]{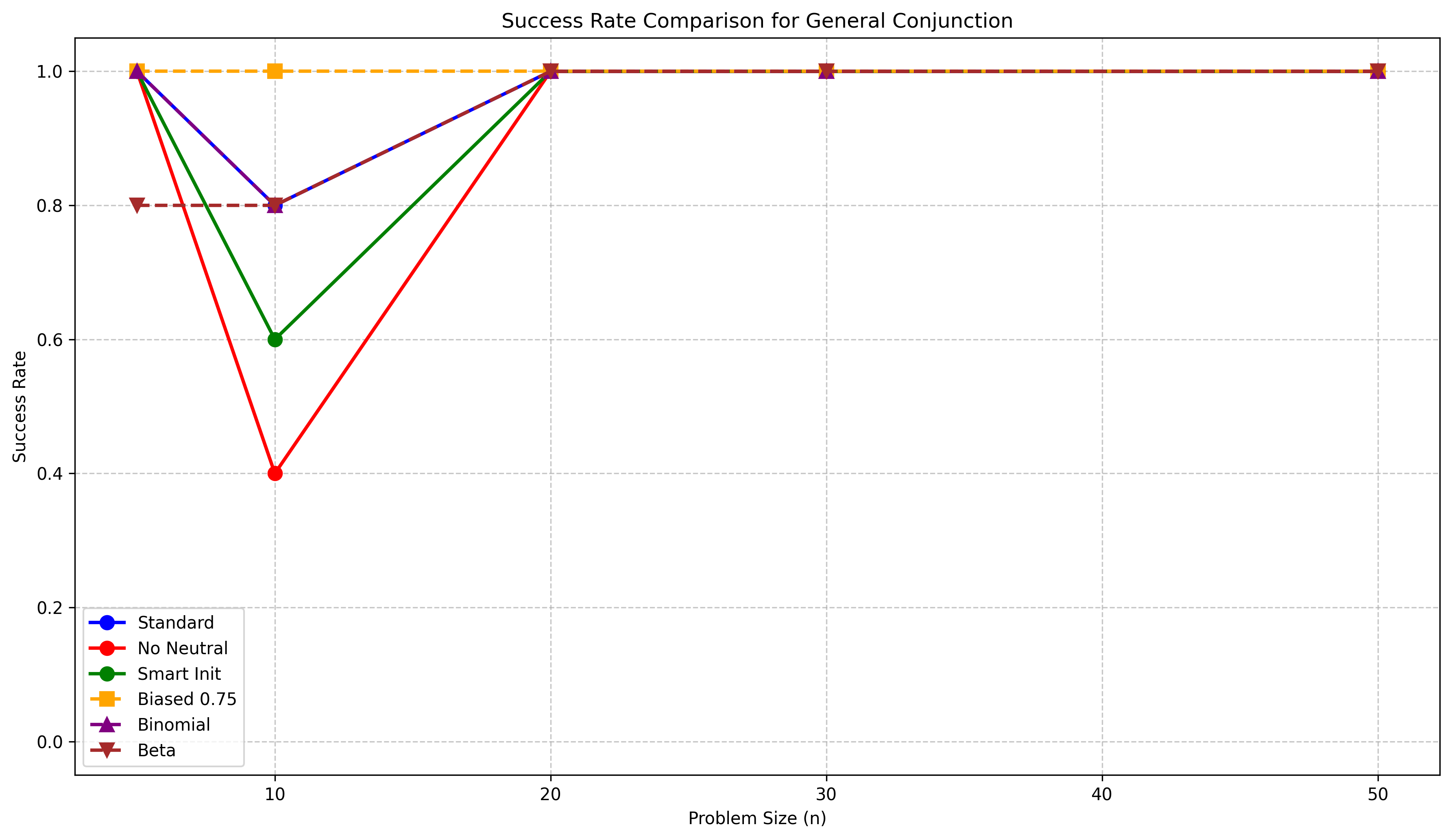}
    \caption{General Conjunction}
  \end{subfigure}
  \hfill
  \begin{subfigure}[b]{0.45\textwidth}
    \includegraphics[width=\textwidth]{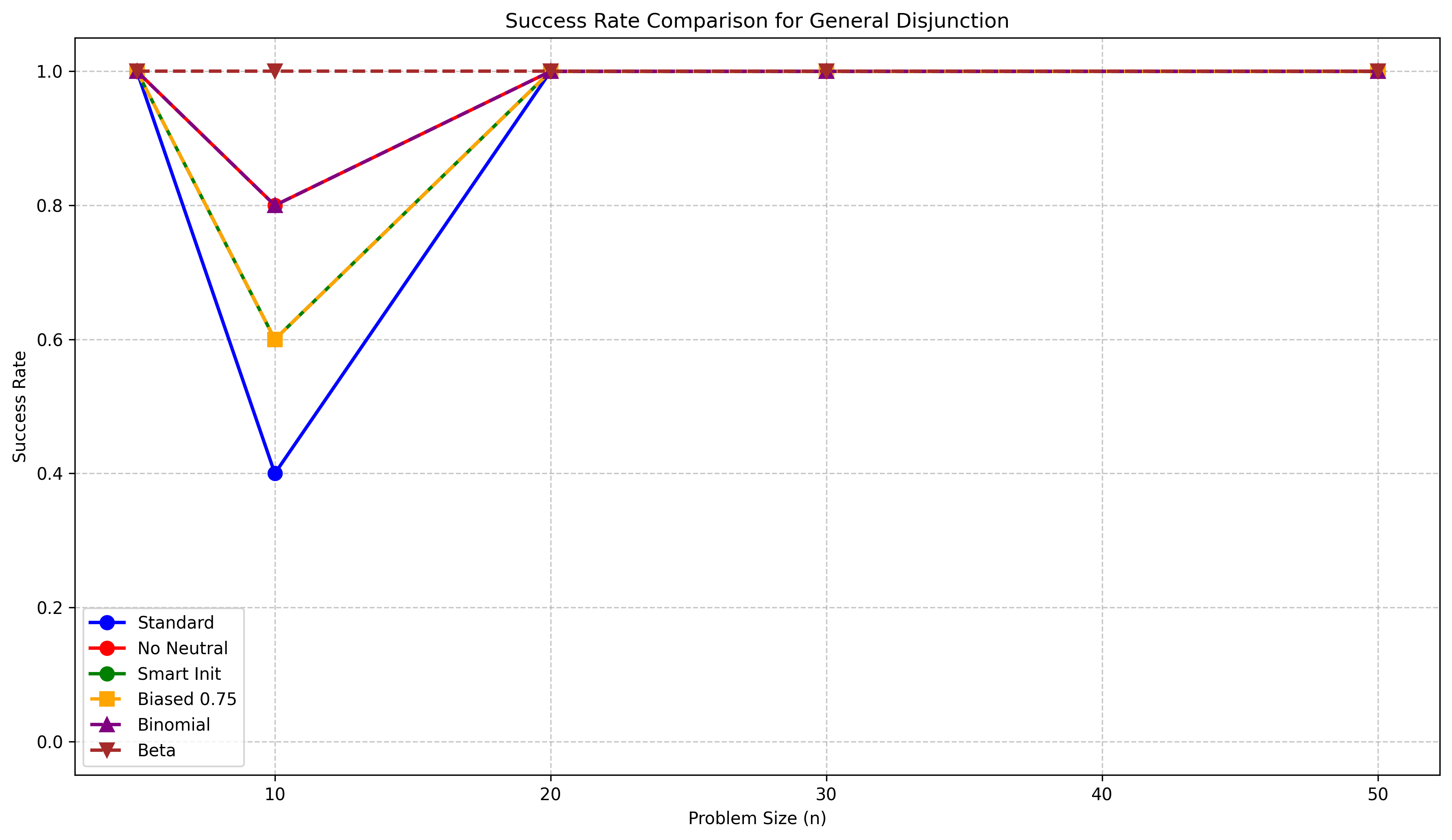}
    \caption{General Disjunction}
  \end{subfigure}

  \vspace{1em}

  \begin{subfigure}[b]{0.45\textwidth}
    \includegraphics[width=\textwidth]{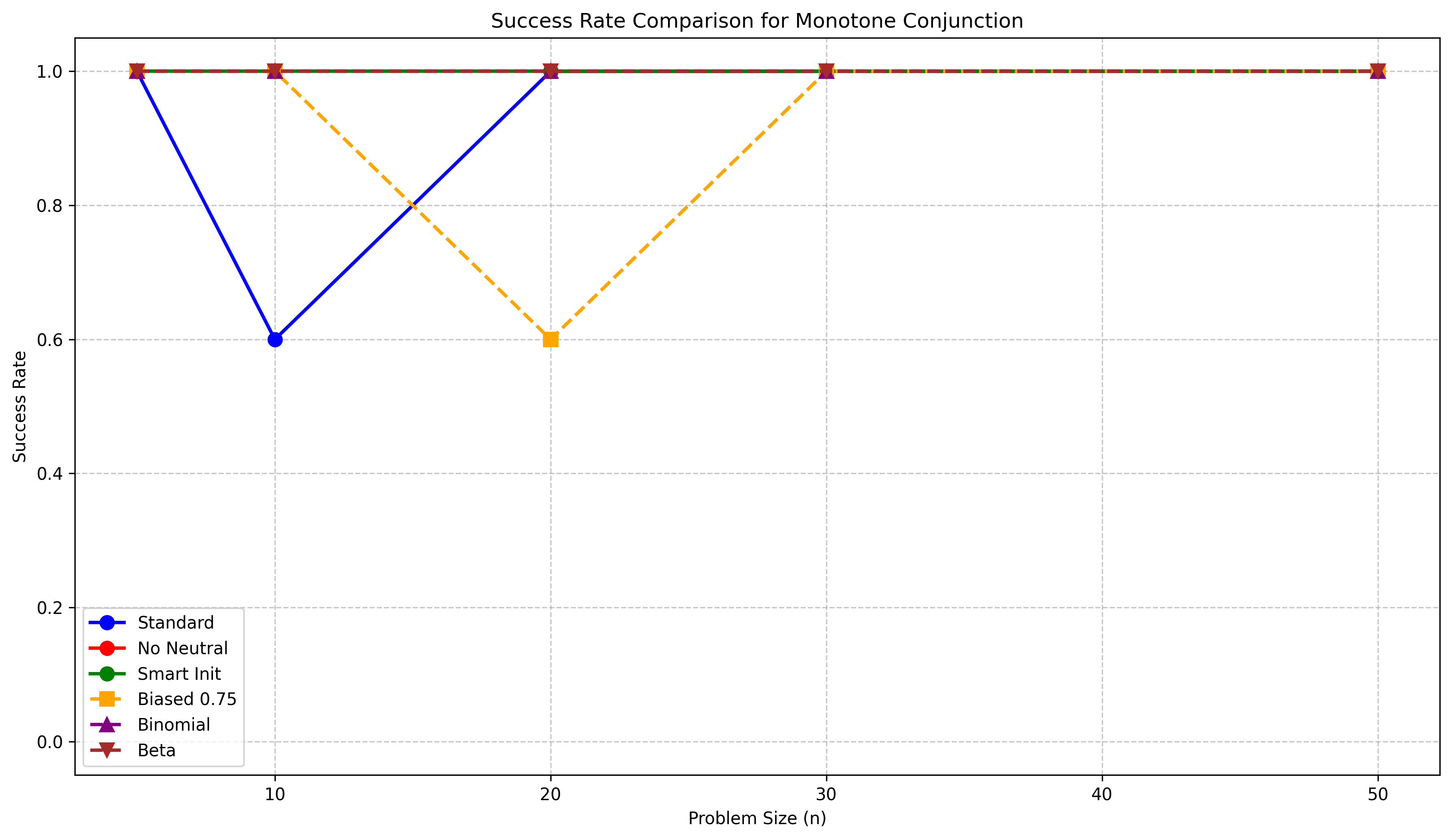}
    \caption{Monotone Conjunction}
  \end{subfigure}
  \hfill
  \begin{subfigure}[b]{0.45\textwidth}
    \includegraphics[width=\textwidth]{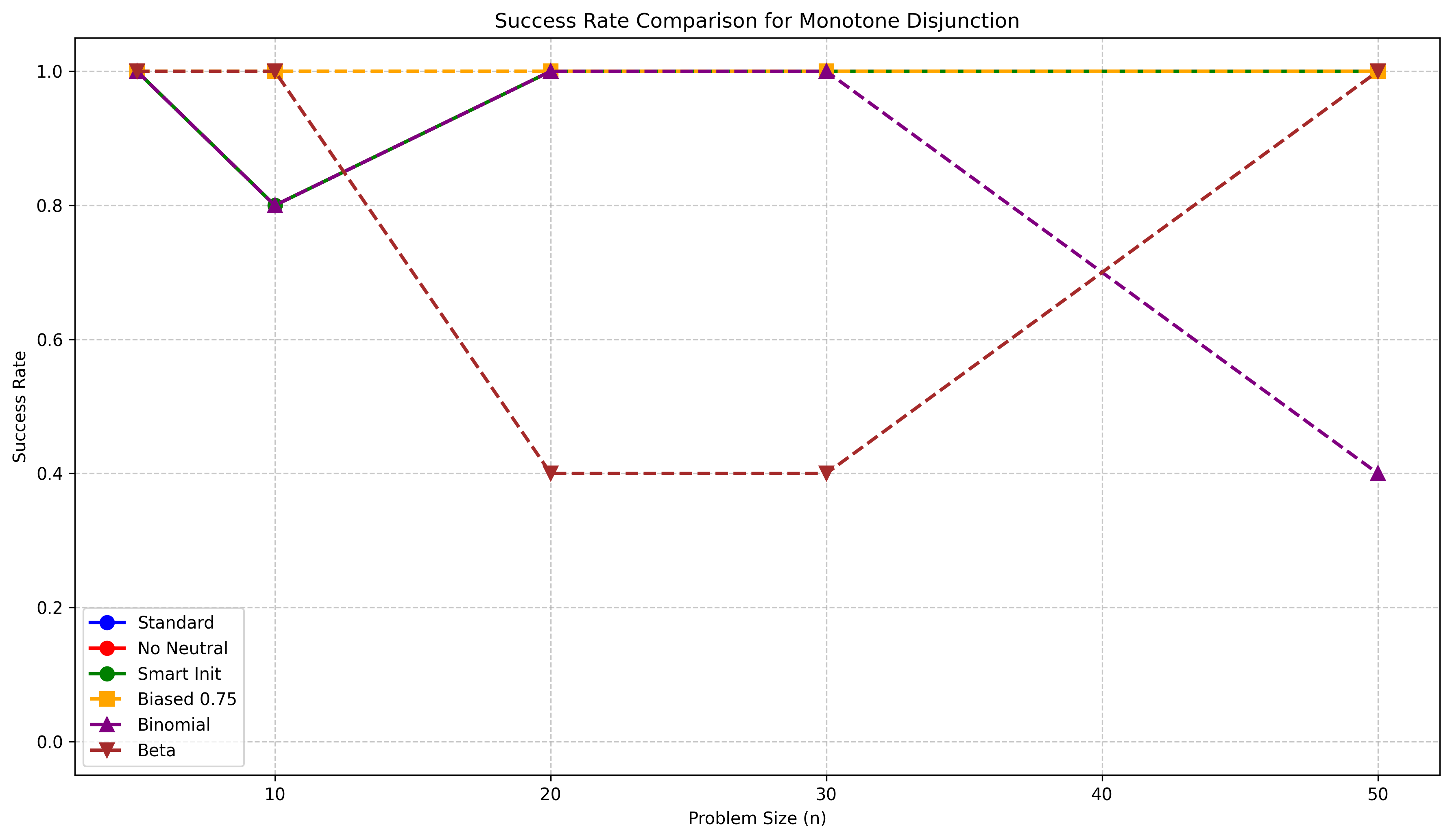}
    \caption{Monotone Disjunction}
  \end{subfigure}

  \vspace{1em}

  \begin{subfigure}[b]{0.45\textwidth}
    \includegraphics[width=\textwidth]{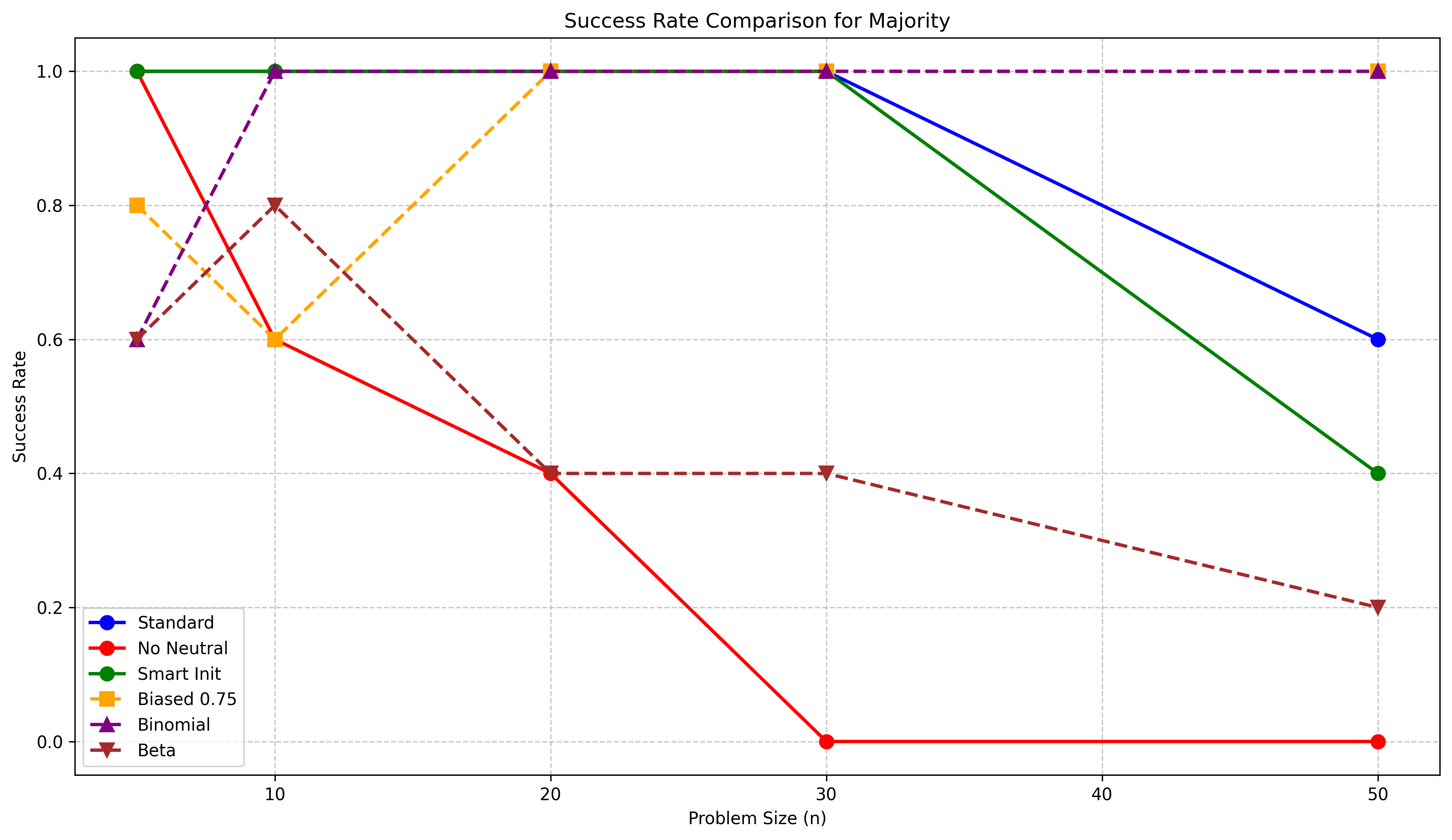}
    \caption{Majority}
  \end{subfigure}
  \hfill
  \begin{subfigure}[b]{0.45\textwidth}
    \includegraphics[width=\textwidth]{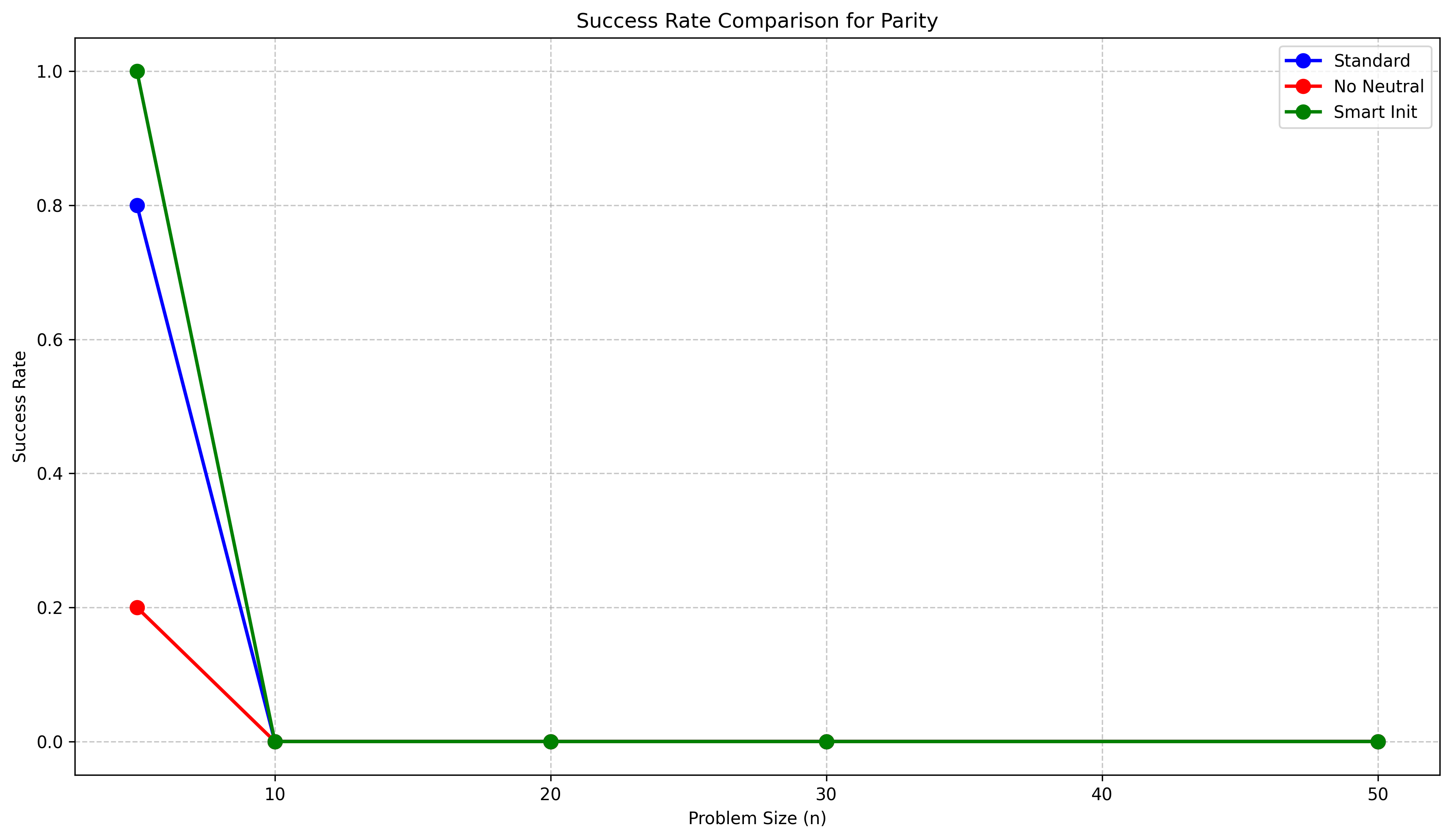}
    \caption{Parity}
  \end{subfigure}

  \caption{Success rates for each function class under Uniform, Binomial, Beta, and Bernoulli(0.75) sampling.}
  \label{fig:dist_perclass}
\end{figure}

To provide class‐specific clarity, Figure~\ref{fig:dist_perclass} arranges six success‐rate plots—one for each function class—across the four regimes in a 3×2 grid.

\FloatBarrier

These per‐class plots reaffirm that general and monotone classes are robust to distributional skew, that Majority becomes easy only under balanced or heavily biased inputs, and that Parity remains intractable under all tested sampling regimes. 

\section{Discussion}

Our empirical investigation confirms and extends the theoretical predictions of Valiant’s evolvability framework while revealing new subtleties in the practical behavior of evolutionary search. Monotone Conjunctions and Disjunctions consistently evolve in a single generation under every regime, validating their provable evolvability. General Conjunctions and Disjunctions, which admit negated literals, succeed almost invariably except at a critical dimension \(n=10\), where success rates and convergence speed degrade sharply (Figures~\ref{fig:standard_summary}–\ref{fig:standard_fitness}). Majority functions occupy an intermediate category—evolvable under balanced or highly biased inputs but failing when neutral drift is disallowed or when sparse inputs obscure informative gradients. Parity remains non‐evolvable throughout, in line with its known non‐SQ learnability.

A key practical bottleneck emerges at \(n=10\) for the general‐class functions. At this dimension the hypothesis‐space size (\(2^{10}\approx10^3\)) is large enough to generate extended neutral plateaus yet too small for neutral drift or random sampling to reliably locate the few beneficial directions needed to escape them. Figure~\ref{fig:standard_summary}d records a peak in neutral mutations exactly at \(n=10\), accompanied by a trough in beneficial moves (Figure~\ref{fig:standard_summary}c) and long stalls in the fitness trajectories (Figure~\ref{fig:standard_fitness}a–b). This “mid‐range plateau” phenomenon suggests that theoretical existence proofs of evolvability may mask significant search‐time challenges when the problem dimension lies in this critical regime.

Neutral mutations prove essential: prohibiting them deepens the \(n=10\) dip for general classes (Figure~\ref{fig:no_neutral_summary}), and causes Majority to collapse entirely at moderate \(n\). Smart initialization substantially reduces generation counts but does not alter the fundamental evolvability thresholds (Figure~\ref{fig:smart_summary}). Distributional skew further modulates evolvability: Majority becomes trivial under Binomial or Bernoulli(0.75) but fails under Beta(2,5) (Figure~\ref{fig:dist_summary}), underscoring that evolvability depends on the interaction between representation, mutation operators, and environmental distribution.

Our study is constrained by computational resources and design choices. First, we limited the input dimension to \(n\le50\); scaling beyond this would require significant compute time or parallel infrastructure. Second, we used a single, fixed neighborhood design and mutation rate; alternative operators (e.g., crossover, multi-bit flips) might yield different dynamics. Third, our trials per configuration (30 runs for Standard, 5 for others) balance statistical confidence with runtime, but larger sample sizes would improve robustness. Fourth, fixed sample sizes (\(s=1000\)) and tolerance (\(\epsilon=0.05\)) simplify comparisons but may not be optimal for all classes or distributions. Finally, the majority neighborhood design—restricting to subsets of size \(k=10\)—is a practical compromise that may not capture all nuances of threshold functions.

These limitations point to several promising directions. On the empirical side, leveraging high‐performance computing or cloud resources would enable exploration of larger \(n\), richer mutation operators, and more exhaustive trial counts. Adaptive strategies—dynamic sample sizes, tolerance schedules, or mutation step‐sizes—could mitigate plateau effects. Extending our simulation to continuous functions or neural architectures would bridge to modern neuroevolution research, testing whether aggregate‐only feedback suffices in complex domains.

Crucially, our findings can guide new theoretical work. The pronounced plateau at \(n=10\) suggests a threshold dimension where formal evolvability bounds may transition from polynomial to super-polynomial time. Empirical neutral‐to-beneficial mutation ratios and fitness‐curve shapes provide concrete parameters for refined theoretical models. In particular, the mixed success of Majority across regimes invites rigorous proof or refutation of its evolvability under Valiant’s model. By identifying distributional conditions and mutation tolerances that enable or preclude evolution, our results lay the groundwork for proofs that characterize exactly when and why Majority is evolvable.

In summary, our empirical framework not only validates known theoretical results but also exposes practical and theoretical frontiers—mid-range plateaus, neutral-drift dependence, and distributional sensitivity—that warrant deeper analysis. Bridging these empirical insights with formal proofs will advance our understanding of evolution as a computational learning algorithm and clarify the true boundaries of evolvability.  

\bibliographystyle{plain}
\bibliography{references}

\end{document}